\documentclass[aps,nofootinbib,showpacs,showkeys,preprint
] {revtex4}

\usepackage{epsf,epsfig,subfigure,axodraw,graphicx,amsmath,amssymb}
\usepackage{color}

\newcommand{\hfive}{${\bf 5}'$}
\newcommand{\hfiveb}{${\bf\overline 5}'$}
\newcommand{\hten}{${\bf 10}'$}
\newcommand{\htenb}{$\overline{\bf 10}'$}
\newcommand{\five}{${\bf 5}$}
\newcommand{\fiveb}{${\bf\overline 5}$}
\newcommand{\ten}{${\bf 10}$}
\newcommand{\tenb}{$\overline{\bf 10}$}

 \def\Z{{\bf Z}}

 \def\Z{{\bf Z}}

\def\fivebt{\overline{\bf 5}}
\def\fivet{{\bf 5}}
\def\tenbt{\overline{\bf 10}}
\def\tent{{\bf 10}}

\begin{document}

\begin{flushright}
{\tt 
~~~~SNUTP 09-005 \quad
\\}
\end{flushright}

\title{\Large\bf SU(5)$_{\rm flip}\times$ SU(5)$'$ from $\Z_{12-I}$}

\author{Ji-Haeng Huh\email{jhhuh@phya.snu.ac.kr},
Jihn E. Kim\email{jekim@ctp.snu.ac.kr}, and Bumseok
Kyae\email{bkyae@kias.re.kr}} \affiliation{Department of Physics
and Astronomy and Center for Theoretical Physics, Seoul National
University, Seoul 151-747, Korea
 }

\begin{abstract}
Based on the $\Z_{12-I}$ orbifold compactification of the
heterotic string theory, we construct a flipped-SU(5) model with
three families of the standard model matter and ingredients for dynamical supersymmetry breaking. The doublet-triplet splittings in the Higgs representations \five$_{-2}$ and \fiveb$_{2}$ are achieved by the couplings \ten$_{1}\cdot$\ten$_{1}\cdot$\five$_{-2}$ and
\tenb$_{-1}\cdot$\tenb$_{-1}\cdot$\fiveb$_{2}$, where \ten$_{1}$
and \tenb$_{-1}$ develop GUT scale vacuum expectation values, breaking the flipped-SU(5) down to the standard model gauge group.
In this model, all the exotic states are decoupled from the low energy physics, and ${\rm sin}^2\theta_W^0=\frac38$.
Above the compactification scale, the flipped-SU(5) gauge symmetry is enhanced to the SO(10) gauge symmetry by including the Kaluza-Klein (KK) modes. The hidden sector gauge group is SU(5)$'$. The threshold correction by the KK modes allow a very wide range for the hidden sector confining scale ($10^{11}$ GeV -- $10^{16}$ GeV).  One family of hidden matter (\tenb$'$ and \five$'$) gives rise to dynamical supersymmetry breaking.

\end{abstract}


\maketitle

\section{Introduction}

The flipped-SU(5) [$\equiv$ SU(5)$\times$U(1)$_X$] model, called SU(5)$_{\rm flip}$, was contrived for the alternative embedding of the standard model (SM) SU(2) singlets in the irreducible representations of the SU(5) grand unified theory (GUT)
\cite{Barr82,Derendinger84,D/Tsplitting,BaeHuh09} in contrast to the
well-known Georgi-Glashow SU(5)$_{\rm GG}$ \cite{GGSU5}.
As a result, a distinctive feature of the SU(5)$_{\rm flip}$ is an
interesting GUT breaking mechanism through the Higgs representation \ten\ of SU(5) rather than the adjoint {\bf 24}, reducing the rank
of the SU(5)$_{\rm flip}$  by one unit.
A great virtue of the SU(5)$_{\rm flip}$ is the relative ease of the
doublet/triplet splitting in the Higgs representations, \five\  and \fiveb, through a simple missing partner mechanism \cite{D/Tsplitting}, which is also a result of such an embedding of the SM fermions.
Another characteristic feature of the SU(5)$_{\rm flip}$ is practically the absence of predicted fermion mass relations between quarks and
leptons in contrast to the SU(5)$_{\rm GG}$ GUT.
As in the case of the SM, supersymmetric (SUSY) extension of the
flipped-SU(5) \cite{Derendinger84} achieves also the gauge
coupling unification with the LEP values of coupling constants
\cite{Raby81}, if the normalization of the hypercharge is assumed
to be that of the SO(10) GUT, $\sin^2\theta_W=\frac38$.

With the advent of string constructions of the SUSY GUT models,
and particularly, with the realization of the difficulty in
obtaining adjoint Higgs for GUT breaking in string theory, the GUT
breaking by the Higgs representations \ten$_{1}$ and
\tenb$_{-1}$ in the SU(5)$_{\rm flip}$ became a great advantage.
Earlier string construction obtaining 4-dimensional (4D) flipped-SU(5) GUTs was done in the fermionic construction
\cite{FlippedSU5}. Recently, a realistic model has been obtained
in a ${\bf Z}_{12-I}$ orbifold construction \cite{FlipKimKyae}.

Since mid 1990s, dynamical SUSY breaking (DSB) toward
phenomenological models has been advocated to resolve the SUSY
flavor problem \cite{DineNelson93}. The well-known simple
dynamical SUSY breaking (DSB) representations in the hidden sector
are \hten\ plus \hfiveb\ of SU(5)$'$ \cite{DSBSU5}, and {\bf
16}$'$ of SO(10)$'$ \cite{DSBSO10}. Other hidden sector gauge groups may be possible, but here we concentrate on a simple SU(5)$'$ model with
only one family, i.e. \hten\ plus \hfiveb, because of its
relatively easy realization in heterotic string compactification.
Recent DSB models at unstable vacua are known to be possible with
vector-like representations in the hidden sector \cite{ISS06}, which became popular because of our familiarity with SUSY QCD. For SU($N$)$'$,
the DSB requirement on the number of flavors in the SUSY QCD is
$N+1\le N_f <\frac32 N$. For SU(5)$'$, $N_f=6$ and 7 satisfy this
requirement. The DSB possibility from the heterotic string has
been suggested by one of the authors at the unstable vacuum
\cite{Kim07un} and at the stable vacuum \cite{Kim07st}. In
particular, the minimal supersymmetric standard model (MSSM)
obtained in \cite{Kim07st} with the SU(5)$'$ gauge group and one
family of \htenb~$\oplus$~\hfive\ in the hidden sector has many nice
features such as the R-parity, one pair of Higgs doublets, and
vector-like exotically charged states (exotics); but the bare value of the weak mixing angle is not $\frac38$. The weak mixing angle would be, however, fitted to the observed one with the power-law type threshold effects contributed by the Kaluza-Klein towers \cite{KimKyaeKK08}, if
relatively larger extra dimensions are assumed.

The so-called SUSY GUTs arise in two disguises: one is (usual) 4D
SUSY GUTs such as the Dimopoulos-Georgi model \cite{DimGeo81} and the
flipped-SU(5) \cite{Derendinger84}, and the other GUTs in higher
dimensions ($D>4$) as discussed in \cite{SUSYGUText}. In a 4D SUSY
GUT, the SM gauge group is obtained by spontaneous symmetry breaking
of the GUT, whereas in a higher dimensional GUT it is achieved by the boundary conditions.  String constructions of the MSSM
\cite{SUSYGUTstring,KimKyaeKK08,stringMSSM,ChoiKimbk06,MSSMothers06} actually provided the idea of the higher dimensional SUSY GUT.
In this paper, we will study a 4D SUSY GUT from a string
compactification.  In particular, based on the ${\bf Z}_{12-I}$
orbifold compactification of the heterotic string theory, we will
construct a SUSY model  SU(5)$_{\rm flip}\times$SU(5)$'$, where the first (second) SU(5) indicates the gauge group of the visible (hidden) sector:
The SU(5)$_{\rm flip}$ for the visible sector is broken to
the SM gauge group by the vacuum expectation values (VEVs) of Higgs fields ${\bf 10}_H'+\overline{\bf
10}_H'$, and the SU(5)$'$ in the hidden sector becomes confined at
lower energies, achieving DSB with one family of
(\htenb~$\oplus$~\hfive) \cite{DSBSU5}.\footnote{We use the one hidden sector matter notation as (\htenb~$\oplus$~\hfive) of SU(5)$'$ to
distinguish it from the visible sector matter notation
$(\overline{\bf 10} \oplus {\bf 5})$ of the flipped-SU(5). }
This model yields MSSM fields plus one pair of $({\bf
10}_H'+\overline{\bf 10}_H')$ and one family of
(\htenb~$\oplus$~\hfive) in the hidden sector. All the other states in this construction are vector-like under the flipped-SU(5).
A nice feature of the flipped-SU(5) model we present in this paper is that it gives a bare value $\frac38$ for $\sin^2\theta_W^0$.
Above the compactification scale the visible sector flipped-SU(5) gauge symmetry is enhanced to SO(10) by including the KK modes.

This paper is organized as follows. In Secs. \ref{sec:USfield} and \ref{sec:TSfield}, we will construct a SUSY GUT model  SU(5)$_{\rm flip}\times$SU(5)$'$ and present the massless spectra from the untwisted and twist sectors. In Sec. \ref{sec:SMsinglets}, we will
discuss the Yukawa couplings needed for realization of the MSSM.
Sec. \ref{sec:effChiral}  will be devoted for the discussion of gauge coupling constants. Finally we will conclude in Sec. \ref{sec:Conclusion}.

\section{${\bf Z}_{12-I}$ orbifold model and $U$ sector fields}
\label{sec:USfield}

We employ the $\Z_{12-I}$ orbifold compactification scheme for the
extra 6D space, which preserves $N=1$ SUSY in the non-compact 4D
spacetime \cite{ChoiKimbk06,ZN}. $\Z_{12-I}$ orbifolds are known
to give phenomenologically interesting MSSMs
\cite{FlipKimKyae,ChoiKimbk06,stringMSSM,KimKyaeKK08}.

The $\Z_{12-I}$ orbifold is an SO(8)$\times$SU(3)
lattice, and the Wilson lines $W_3$ and $W_4$ ($=W_3$) can be
introduced in the 2D SU(3) lattice \cite{ZN,ChoiKimbk06}.
We take the following shift vector $V$ and the Wilson line $W_3$,
\begin{equation}
\begin{array}{l}
V=\textstyle\left( 0~ 0~ 0~ 0~ 0~ ;\frac{-1}{6}~\frac{-1}{6}~  \frac{-1}{6}~ \right)\left(0~0~0~0~0~\frac{1}{4}~
\frac{1}{4}~ \frac{-2}{4} \right)' ,\\
W_3=W_4\equiv W=\textstyle\left(
\frac23~\frac23~\frac23~\frac23~\frac23~;
~0~\frac{-2}{3}~\frac{2}{3} \right)\left( \frac23~\frac23~\frac23~\frac23~ 0~\frac{-2}{3}~0~0 \right)'
\end{array}
\end{equation}
which are associated with the boundary conditions of the
left moving bosonic string.
For modular invariance in ${\bf Z}_{12-I}$ orbifold
compactification, $V$ and $W$ should be specially related to the
twist vector $\phi=(\frac{5}{12}~\frac{4}{12}~\frac{1}{12})$,
which is associated with the boundary conditions of the right
moving superstrings, preserving only $N=1$ SUSY in 4D.  The twist
vector $\phi=(\frac{5}{12}~\frac{4}{12}~\frac{1}{12})$ specifies
the ${\bf Z}_{12-I}$ orbifold. This model gives
\begin{eqnarray}
V^2-\phi^2 = \frac16, \quad W^2=\frac{16}{3}, \quad V\cdot W= \frac{-1}{6} .
\end{eqnarray}
Hence, the modular invariance conditions
in ${\bf Z}_{12-I}$ orbifold compactification are satisfied \cite{ChoiKimbk06}: $12\cdot (V^2-\phi^2)=$ even integer, $12\cdot W^2=$ even integer, and $12V\cdot W=$ integer.

The string excited states are irrelevant to low energy
physics. The massless conditions for the left and right moving
strings on the orbifold ${\bf Z}_{12-I}$ are
\begin{eqnarray}
\label{massless} &&{\rm left\ moving\ string}:~\
\frac{(P+kV_f)^2}{2}+\sum_iN^L_i\tilde{\phi}_i -\tilde c_k=0 ,
\nonumber \\
&&{\rm right\ moving\ string}:\
\frac{(s+k\phi)^2}{2}+\sum_iN^R_i\tilde{\phi}_i-c_k=0, \nonumber
\end{eqnarray}
where $k=0,1,2,\cdots,11$, $V_f=(V+m_fW)$ with $m_f=0,+1,-1$, and
$i$ runs over $\{1,2,3,\bar{1},\bar{2},\bar{3}\}$.   Here
$\tilde{\phi}_j\equiv k\phi_j$ mod $Z$ such that
$0<\tilde{\phi}_j\leq 1$, and $\tilde{\phi}_{\bar{j}}\equiv
-k\phi_j$ mod Z such that $0<\tilde{\phi}_{\bar{j}}\leq 1$.
$N^L_i$ and $N^R_i$ indicate oscillating numbers for the left and
right movers. $P$ and $s$ [$\equiv (s_0,\tilde{s})$] are the ${\rm
E_8\times E_8'}$ and SO(8) weight vectors, respectively. The
values of $\tilde{c}_k$, $c_k$ are found in
Refs.~\cite{ChoiKimbk06,ZN,FlippedSU5}.

The multiplicity for a given massless state is
calculated with the GSO projector in the ${\bf Z}_{12-I}$
orbifold,
\begin{eqnarray} \label{phase}
{\mathcal P}_k(f) = \frac{1}{12\cdot 3}\sum_{l = 0 }^{11}
\tilde{\chi} ( \theta^k , \theta^l ) e^{2 \pi i l\Theta_k} ,
\nonumber
\end{eqnarray}
where $f$ $(=\{f_0, f_+, f_-\})$ denotes twist sectors associated
with $kV_f=kV$, $k(V+W)$, $k(V-W)$. The phase $\Theta_k$ is given
by
\begin{eqnarray}
&&\textstyle \Theta_k = \sum_i (N^L_i - N^R_i ) \hat{\phi}_i
+ (P+\frac{k}{2}V_f) V_f
 -(\tilde{s} + \frac{k}{2} \phi) \phi ,
 \nonumber
\end{eqnarray}
where $\hat{\phi}_j = \phi_{j}$ and
$\hat{\phi}_{\bar{j}}=-\phi_j$. Here, $\tilde{\chi}(\theta^k,
\theta^l)$ is the degeneracy factor summarized in
Ref.~\cite{ChoiKimbk06,ZN,FlippedSU5}. Note that ${\cal
P}_k(f_0)={\cal P}_k(f_+)={\cal P}_k(f_-)$ for $k=0,3,6,9$.
In addition, the left moving states should satisfy
\begin{eqnarray}\label{condi3}
P\cdot W =0~~{\rm mod}~~{\rm Z}~~{\rm
in~the}~U,~T_3,~T_6,~T_9~{\rm sectors}. \nonumber
\end{eqnarray}

The massless gauge sector corresponds to the states
satisfying $P\cdot V=$ integer, and $P\cdot W=$ integer. They are
\begin{eqnarray} \label{su5root}
&&{\rm SU(5)} ~:~~ \textstyle\left(\underline{1~-1~0~0~0~};0^3\right)(0^8)'
\\
&&{\rm SU(5)}' ~:~ \left\{
\begin{array}{l}\textstyle(0^8)\left(
\underline{1~-1~0~0}~0~0~0~0\right)'  \label{su5'root}
\\
\textstyle\pm (0^8)\left(\underline{+---}-+++\right)'
\end{array}
\right.
\\
&&{\rm SU(2)}' ~:~~\pm(0^8)\left(++++++++\right)' ,
\label{su2'root}
\end{eqnarray}
where the underline means all possible permutations.
Thus, the gauge group is
\begin{eqnarray}
{\rm \left[\{SU(5)\times U(1)_X\}\times
 U(1)^3\right]\times\left[SU(5)\times
SU(2) \times U(1)^3\right]'},
\end{eqnarray}
where SU(5)$\times$U(1)$_X$ is identified with the flipped SU(5). The U(1)$_X$ charge operator of the flipped-SU(5) is \cite{FlipKimKyae},
\begin{eqnarray} \label{U(1)X}
X= \textstyle \frac{1}{\sqrt{40}}\left( -2~ -2~ -2~ -2~ -2~ ;~0^3 \right)\left(0^8 \right)' .
\end{eqnarray}
The normailization factor $\frac{1}{\sqrt{40}}$ is determined such
that the norm of the $X$ (in general all U(1) charge operators in
the level one heterotic string theory \cite{ChoiKimbk06}) is
$\frac{1}{\sqrt{2}}$. This value is exactly the one given as the
normalization required for the SU(5)$\times$U(1)$_X$ embedded in
SO(10).
Since the standard model hypercharge is defined as
\begin{eqnarray}
Y= \textstyle \sqrt{\frac{3}{5}}\left( \frac13~ \frac13~ \frac13~
\frac12~ \frac12~ ;~0^3 \right)\left(0^8 \right)' ,
\end{eqnarray}
the weak mixing angle at the string scale is
$\sin^2\theta_W^0=\frac38 $.  From now on, we will drop the
normalization factor ``$\frac{1}{\sqrt{40}}$'' and  ``$\sqrt{\frac{3}{5}}$'' just for simplicity.

\begin{table} [h]
\begin{center}
\begin{tabular}{|c|c|c||c|}
\hline  Visible States & $P\cdot V$ & $~\chi~$ & ~SU(5)$_X$~\\
\hline  $(\underline{+----};+--)(0^8)'$ & $\frac{1}{12}$ & L & ${\bf 5}_{3}$\\
 $(\underline{+++--};--+)(0^8)'$ & $\frac{1}{12}$ & L & $\overline{\bf 10}_{-1}$\\
 $(+++++;-+-)(0^8)'$ & $\frac{1}{12}$ & L & ${\bf 1}_{-5}$\\
\hline
\end{tabular}
\end{center}
\caption{The $U$ sector chiral states. There is
no hidden sector chiral states and no flipped-SU(5) singlets.
}\label{tb:untwistedvi}
\end{table}

The massless chiral matter in the $U$ sector
($U$) are the states satisfying $P\cdot V=$ $\{\frac{-5}{12}$ or
$\frac{4}{12}$ or $\frac{1}{12}\}$, and $P\cdot W =$ integer.
In Table \ref{tb:untwistedvi}, the chiral fields in the $U$
sector are tabulated. Note that there does not appear any flipped-SU(5)
singlets in $U$. From the $U$ sector, we obtain one family of the MSSM matter
\begin{eqnarray}
\overline{\bf 10}_{-1}+{\bf 5}_{3}+{\bf 1}_{-5} , ~~{\rm (and~
their~{\cal CTP}~conjugates)} ,
\end{eqnarray}
where $\overline{\bf 10}_{-1}$, ${\bf 5}_{3}$, ${\bf
1}_{-5}$ contain $\{d^c_L,q_L,\nu^c_L\}$, $\{u^c_L,l_L\}$, and
$e^c_L$, respectively. It is tempting to interpret this as the third (top quark) family, but the low dimensional Yukawa couplings prefer one in the twisted sector as the third family.

\section{Twisted sector fields}\label{sec:TSfield}

There are 11 twisted sectors, $T_k$ with $k=1,2,\cdots,11$. The ${\cal CTP}$ conjugates of the chiral states in $T_k$ is provided in $T_{12-k}$. Thus, it is sufficient to consider $k=1,2,\cdots,6$.
While the $U$ and $T_6$ sectors contain both chiral
states and their ${\cal CTP}$ conjugates, $T_1$, $T_2$, $T_4$, and
$T_7$ ($T_{11}$, $T_{10}$, $T_8$, and $T_5$) sectors yield only the
left-handed (right-handed) chiral states. The $T_3$ sector
includes both left- and right-handed chiral states. So we will
take ${\cal CTP}$ conjugations for the right-handed states from
the $T_3$ and $T_5$ sectors.

\subsection{The flipped-SU(5) spectrum}

The visible sector chiral states of the twisted sectors are
\begin{eqnarray}
&& T_4:~2(\overline{\bf 10}_{-1}+{\bf 5}_{3}+{\bf 1}_{-5}),\ 2({\bf 5}_{-2}+\overline{\bf 5}_{2}),\label{EqObs1}\\
&& T_3, T_9:~ ({\bf 10}_{1}+\overline{\bf 10}_{-1}),
\label{EqObs2}\\
&& T_7:~ ({\bf 5}_{-2}+\overline{\bf 5}_{2}),\label{EqObs3}\\
&& T_6:~  3({\bf 5}_{-2}+\overline{\bf 5}_{2}).
\label{EqObs4}
\end{eqnarray}
To get the left-handed states from the $T_9$ and $T_7$ sectors, we acted the ${\cal CTP}$ conjugations to the right-handed states of $T_3$ and $T_5$ sectors.   From Table \ref{tb:T4} (or Eq.~(\ref{EqObs1})), we note that two families of the MSSM matter fields appear from $T_4$. Together
with one family from the $U$ sector, thus, they form a three
family model, including the three right-handed neutrinos.

\begin{table}[h]
{\tiny
\begin{center}
\begin{tabular}{|c|c|c|c||c|c|}
\hline  $P+4V$ & $~\chi~$ & $(N^L)_j$ &
${\cal P}_4(f_0)$ & ~SU(5)$_X$~\\
\hline
$\left(\underline{+----}~;\frac{-1}{6}~\frac{-1}{6}~
\frac{-1}{6}\right)(0^8)'$ & L & $0$ &  $2$ &$2\cdot
{\bf 5}_{3}$ \\
$\left(\underline{+++--}~;\frac{-1}{6}~\frac{-1}{6}~
\frac{-1}{6}\right)(0^8)'$ & L & $0$ &  $2$ &$2\cdot \overline{\bf 10}_{-1}$  \\
$\left(+++++~;\frac{-1}{6}~\frac{-1}{6}~\frac{-1}{6} \right)
(0^8)'$ & L & $0$ & $2$ &$2\cdot {\bf 1}_{-5}$  \\ \hline
$\left(\underline{1~0~0~0~0}~;\frac{1}{3}~\frac{1}{3}~
\frac{1}{3}\right)(0^8)'$ & L & $0$ & $2$ & $2\cdot {\bf 5}_{-2}$  \\
$\left(\underline{-1~0~0~0~0}~;\frac{1}{3}~\frac{1}{3}~
\frac{1}{3}\right)(0^8)'$ & L & $0$ & $2$ & $2\cdot \overline{\bf 5}_{2}$ \\
\hline
$\left(0~0~0~0~0~;\frac{-2}{3}~\frac{-2}{3}~\frac{-2}{3}
\right)(0^8)'$ & L & $0$ & $3$ &$3\cdot {\bf 1_0}$  \\
$\left(0~0~0~0~0~;\frac{-2}{3}~\frac{1}{3}~\frac{1}{3}
\right)(0^8)'$ & L & $1_{\bar 1},1_{2},1_{3}$ & $2,3,2$ &$(2+3+2)\cdot {\bf 1_0}$  \\
$\left(0~0~0~0~0~;\frac{1}{3}~\frac{-2}{3}~\frac{1}{3}
\right)(0^8)'$ & L & $1_{\bar 1},1_{2},1_{3}$ & $2,3,2$ &$(2+3+2)\cdot {\bf 1_0}$  \\
$\left(0~0~0~0~0~;\frac{1}{3}~\frac{1}{3}~\frac{-2}{3}
\right)(0^8)'$ & L & $1_{\bar 1},1_{2},1_{3}$ & $2,3,2$ &$(2+3+2)\cdot {\bf 1_0}$  \\
\hline \hline
 $P+4V_+$ & $~\chi~$ & $(N^L)_j$ & ${\cal
P}_4(f_+)$ & ~(SU(5)$_X$; SU(5)$'$, SU(2)$'$)~
  \\
\hline $\left(\frac16~\frac16~\frac16~\frac16~\frac16~;
\frac{-1}{6}~\frac16~\frac12\right)
\left(\underline{\frac23~\frac{-1}{3}~\frac{-1}{3}
~\frac{-1}{3}}~0~\frac13~0~0\right)'$ & L & $0$ &
 $3$ & $3\cdot ({\bf 1}_{-5/3}; {\bf 5}',{\bf 1}')$
\\
$\left(\frac16~\frac16~\frac16~\frac16~\frac16~;
\frac{-1}{6}~\frac16~\frac12\right)
\left(\frac16~\frac16~\frac16~\frac16~\frac12~
\frac{-1}{6}~\frac{-1}{2}~\frac{-1}{2}\right)'$ & L & 0 & 3 &
\\ \hline
$\left(\frac16~\frac16~\frac16~\frac16~\frac16~;
\frac{-1}{6}~\frac16~\frac12\right)
\left(\frac{-1}{3}~\frac{-1}{3}~\frac{-1}{3}~\frac{-1}{3}
~0~\frac{-2}{3}~0~0\right)'$ & L & $0$ & $2$ & $2\cdot ({\bf
1}_{-5/3}; {\bf 1}',{\bf 2}')$
\\
$\left(\frac16~\frac16~\frac16~\frac16~\frac16~;
\frac{-1}{6}~\frac16~\frac12\right)
\left(\frac{1}{6}~\frac{1}{6}~\frac{1}{6}~\frac{1}{6}
~\frac12~\frac{-1}{6}~\frac12~\frac12\right)'$ & L & $0$ &$2$ &
\\ \hline
$\left(\frac16~\frac16~\frac16~\frac16~\frac16~;
\frac{-1}{6}~\frac16~\frac12\right)
\left(\frac{1}{6}~\frac{1}{6}~\frac{1}{6}~\frac{1}{6}
~\frac{-1}{2}~\frac{-1}{6}~\frac{-1}{2}~\frac12\right)'$ & L & $0$ &$2$ & $2\cdot {\bf 1}_{-5/3}$
  \\
$\left(\frac16~\frac16~\frac16~\frac16~\frac16~;
\frac{-1}{6}~\frac16~\frac12\right)
\left(\frac{1}{6}~\frac{1}{6}~\frac{1}{6}~\frac{1}{6}
~\frac{-1}{2}~\frac{-1}{6}~\frac12~\frac{-1}{2}\right)'$ & L & $0$ &$2$ & $2\cdot {\bf 1}_{-5/3}$
  \\
\hline \hline
  $P+4V_-$ & $~\chi~$ & $(N^L)_j$ & ${\cal
P}_4(f_-)$ & ~(SU(5)$_X$; SU(5)$'$, SU(2)$'$)~
\\
\hline $\left(\frac{-1}{6}~\frac{-1}{6}~\frac{-1}{6}~\frac{-1}{6}
~\frac{-1}{6};\frac{-1}{6}~\frac{-1}{2}~\frac16\right)
\left(\underline{\frac{-2}{3}~\frac{1}{3}~\frac{1}{3}
~\frac{1}{3}}~0~\frac{-1}{3}~0~0\right)'$ & L & $0$ & $3$ &
$3\cdot ({\bf 1}_{5/3}; \overline{\bf 5}',{\bf 1}')$
\\
$\left(\frac{-1}{6}~\frac{-1}{6}~\frac{-1}{6}~\frac{-1}{6}
~\frac{-1}{6}~;\frac{-1}{6}~\frac{-1}{2}~\frac16\right)
\left(\frac{-1}{6}~\frac{-1}{6}~\frac{-1}{6}~\frac{-1}{6}
~\frac{-1}{2}~\frac{1}{6}~\frac{1}{2}~\frac{1}{2}\right)'$ & L & 0 & 3 &
  \\ \hline
$\left(\frac{-1}{6}~\frac{-1}{6}~\frac{-1}{6}~\frac{-1}{6}
~\frac{-1}{6}~;\frac{-1}{6}~\frac{-1}{2}~\frac16\right)
\left(\frac{1}{3}~\frac{1}{3}~\frac{1}{3}~\frac{1}{3}~0~
\frac{2}{3}~0~0\right)'$ & L & $0$ & $2$ & $2\cdot ({\bf
1}_{5/3};{\bf 1}', {\bf 2}')$
\\
$\left(\frac{-1}{6}~\frac{-1}{6}~\frac{-1}{6}~\frac{-1}{6}
~\frac{-1}{6}~;\frac{-1}{6}~\frac{-1}{2}~\frac16\right)
\left(\frac{-1}{6}~\frac{-1}{6}~\frac{-1}{6}~\frac{-1}{6}
~\frac{-1}{2}~\frac{1}{6}~\frac{-1}{2}~\frac{-1}{2} \right)'$ & L
& $0$ &$2$ &
\\ \hline
$\left(\frac{-1}{6}~\frac{-1}{6}~\frac{-1}{6}~\frac{-1}{6}
~\frac{-1}{6}~;\frac{-1}{6}~\frac{-1}{2}~\frac16\right)
\left(\frac{-1}{6}~\frac{-1}{6}~\frac{-1}{6}~\frac{-1}{6}
~\frac{1}{2}~\frac{1}{6}~\frac{1}{2}~\frac{-1}{2}\right)'$ & L &
$0$ &$2$ & $2\cdot {\bf 1}_{5/3}$
\\
$\left(\frac{-1}{6}~\frac{-1}{6}~\frac{-1}{6}~\frac{-1}{6}
~\frac{-1}{6}~;\frac{-1}{6}~\frac{-1}{2}~\frac16\right)
\left(\frac{-1}{6}~\frac{-1}{6}~\frac{-1}{6}~\frac{-1}{6}
~\frac{1}{2}~\frac{1}{6}~\frac{-1}{2}~\frac12\right)'$ & L & $0$ & $2$ & $2\cdot {\bf 1}_{5/3}$
  \\
\hline
\end{tabular}
\end{center}
\caption{Chiral matter  states in the $T_4^0$, $T_4^+$, and $T_4^-$ sectors. The multiplicities are shown as the coefficient in the last column.}\label{tb:T4}
 }
\end{table}
In Table \ref{tb:T6} some Higgs doublets are shown. Altogether, there appear six pairs of Higgs doublets from $T_4,T_7$ and $T_6$,
among which therefore the candidates of the MSSM Higgs doublets
are chosen. We will explain in Sec. \ref{sec:SMsinglets} that except one pair of $\{{\bf 5}_{-2},\overline{\bf 5}_{2}\}$, the other pairs of five-plets with $X=\pm 2$ in the $T_4$, $T_5$, and $T_6$ sectors achieve
superheavy masses, when some singlets under
$[$SU(5)$\times$U(1)$_X]\times$[SU(5)$\times$ SU(2)]$'$ obtain
VEVs of order the string scale. We regard the remaining one pair of
$\{{\bf 5}_{-2},\overline{\bf 5}_{2}\}$ as the Higgs containing
the MSSM Higgs. We will explain also how to decouple the triplets appearing in such five-plets in Sec. \ref{sec:SMsinglets}.

\begin{table}[h]
{\tiny
\begin{center}
\begin{tabular}{|c|c|c|c|c||c|}
\hline  $P+6V$ & $~\chi~$ & $(N^L)_j$ & $\Theta_L$ & ${\cal P}_6$
&~SU(5)$_X$~
\\
\hline
 $\left(\underline{1~0~0~0~0}~;0~0~0\right)
 \left(0~0~0~0~0~\frac{-1}{2}~\frac{1}{2}~0 \right)'$ &
L & $0$ & $\frac{-1}{3}$ & $3$ & $3\cdot {\bf 5}_{-2}$
\\
 $\left(\underline{-1~0~0~0~0}~;0~0~0\right)
 \left(0~0~0~0~0~\frac{1}{2}~\frac{-1}{2}~0 \right)'$ &
L & $0$ & $\frac{-1}{3}$ & $3$ & $3\cdot \overline{\bf 5}_{2}$
\\ \hline
 $\left(0~0~0~0~0~;0~ 1~0\right)
 \left(0~0~0~0~0~\frac{1}{2}~\frac{-1}{2}~0 \right)'$ &
L & $0$ & $\frac{-1}{2}$ & $2$ &  $2\cdot {\bf 1}_{0}$
\\
 $\left(0~0~0~0~0~;0~0 ~1\right)
 \left(0~0~0~0~0~\frac{-1}{2}~\frac{1}{2}~0 \right)'$ &
L & $0$ & $\frac{-1}{2}$ & $2$ & $2\cdot {\bf 1}_{0}$
\\
 $\left(0~0~0~0~0~;0~ -1~0\right)
 \left(0~0~0~0~0~\frac{-1}{2}~\frac{1}{2}~0 \right)'$ &
L & $0$ & $\frac{-1}{6}$ & $2$ & $2\cdot {\bf 1}_{0}$
\\
 $\left(0~0~0~0~0~;0~0 -1\right)
 \left(0~0~0~0~0~\frac{1}{2}~\frac{-1}{2}~0 \right)'$ &
L & $0$ & $\frac{-1}{6}$ & $2$ & $2\cdot {\bf 1}_{0}$
\\
\hline
\end{tabular}
\end{center}
\caption{Massless states satisfying $P\cdot W=0 $ mod $Z$ in
$T_6$. } \label{tb:T6} }
\end{table}

To break the flipped-SU(5) down to the SM, we need \ten$_1$ ($\equiv{\bf 10}_{H}$) and \tenb$_{-1}$ ($\equiv\overline{\bf 10}_{H}$), which appear from $T_3$ and $T_9$ as shown in Table \ref{tb:T3states}. As explained later, they couple to the $\{{\bf 5}_{-2},\overline{\bf 5}_{2}\}$ ($\equiv\{{\bf 5}_{h},\overline{\bf 5}_{h}\}$) so that the pseudo-Goldstone mode $\{D,D^c\}$ included in $\{{\bf 10}_{H}, \overline{\bf 10}_{H}\}$ pair up with the triplets
contained in $\{{\bf 5}_{-2},\overline{\bf 5}_{2}\}$ to be
superheavy.
\begin{table}[h]
{\tiny
\begin{center}
\begin{tabular}{|c|c|c|c|c||c|}
\hline  $P+3V$ & $~\chi~$ & $(N^L)_j$ & $\Theta_{L,R}$ & ${\cal P}_3$& ~SU(5)$_X$~
\\
\hline
 $\left(\underline{+++--}~;0~0~0 \right)
 \left(0~0~0~0~0~\frac{-1}{4}~\frac{-1}{4}~\frac{2}{4}
 \right)'$ &L & $0$ & $\frac{-1}{3}$ & $1$ & $1\cdot \overline{\bf 10}_{-1}$
\\
 $\left(\underline{+++--}~;0~0~0 \right)
 \left(0~0~0~0~0~\frac{-1}{4}~\frac{-1}{4}~\frac{2}{4}
 \right)'$ &R & $0$ & $\frac{-2}{3}$ & $1$ &  $1\cdot {\bf 10}_{1}^{~*}$ , or\\
 &L&&& & ($1\cdot {\bf 10}_{1}$ from $T_9$)
\\ \hline
 $\left(0~0~0~0~0~;\frac{-1}{2}~\frac{-1}{2}~\frac{-1}{2}
 \right)\left(0~0~0~0~0~\frac{3}{4} ~\frac{-1}{4}~ \frac{-1}{2}\right)'$ & L & $0$ & $\frac23$ & $1$ &  $1\cdot {\bf 1}_0$
\\
 $\left(0~0~0~0~0~;\frac{-1}{2}~\frac{-1}{2}~\frac{-1}{2}
 \right)\left(0~0~0~0~0~\frac{3}{4} ~\frac{-1}{4}~ \frac{-1}{2}\right)'$ & R & $0$ & $\frac13$ & $1$ &  $1\cdot {\bf 1}_0^{~*}$
\\
 $\left(0~0~0~0~0~;\frac{-1}{2}~\frac{1}{2}~\frac{1}{2}
 \right)\left(0~0~0~0~0~\frac{3}{4} ~\frac{-1}{4}~ \frac{-1}{2}\right)'$ &L & $0$ & $\frac13$ & $1$ &  $1\cdot {\bf 1}_0$
\\
 $\left(0~0~0~0~0~;\frac{-1}{2}~\frac{1}{2}~\frac{1}{2}
 \right)\left(0~0~0~0~0~\frac{3}{4} ~\frac{-1}{4}~ \frac{-1}{2}\right)'$ & R & $0$ & $0$ & $2$ &  $2\cdot {\bf 1}_0^{~*}$
\\
 $\left(0~0~0~0~0~;\frac{1}{2}~\frac{1}{2}~\frac{-1}{2}
 \right)\left(0~0~0~0~0~\frac{-1}{4} ~\frac{3}{4}~ \frac{-1}{2}\right)'$ &L & $0$ & $\frac13$ & $1$ &  $1\cdot {\bf 1}_0$
\\
 $\left(0~0~0~0~0~;\frac{1}{2}~\frac{1}{2}~\frac{-1}{2}
 \right)\left(0~0~0~0~0~\frac{-1}{4} ~\frac{3}{4}~ \frac{-1}{2}\right)'$ &
R & $0$ & $0$ & $2$ &  $2\cdot {\bf 1}_0^{~*}$
\\
 $\left(0~0~0~0~0~;\frac{1}{2}~\frac{1}{2}~\frac{-1}{2}
 \right)\left(0~0~0~0~0~\frac{-1}{4} ~\frac{-1}{4}~ \frac{1}{2}\right)'$ &L & $1_{1},1_{3}$ & $0$,$\frac{-1}{3}$ & $2$,$1$ &  $(2+1)\cdot {\bf 1}_0$
\\
 $\left(0~0~0~0~0~;\frac{1}{2}~\frac{1}{2}~\frac{-1}{2}
 \right)\left(0~0~0~0~0~\frac{-1}{4} ~\frac{-1}{4}~ \frac{1}{2}\right)'$ &R & $1_{1}$,$1_{3}$ & $\frac{-1}{3}$,$\frac{-2}{3}$ & $1$,$1$ &$(1+1)\cdot {\bf 1}_0^{~*}$
\\
\hline
\end{tabular}
\end{center}
\caption{Massless states from $T_3$. The starred chirality R  states in $T_3$ can be represented
also by un-starred chirality L  states with the opposite quantum numbers in $T_9$.} \label{tb:T3states}
}
\end{table}

\subsection{The hidden-sector SU(5)$'$ spectrum}
The hidden sector fields appear from twisted sectors. The chiral multiplets under SU(5)$'\times$SU(2)$'$
are listed as follows.
\begin{eqnarray}
&& T_4:~3({\bf 5}',{\bf 1}')_{-5/3},~ 3(\overline{\bf 5}',{\bf 1}')_{5/3},~ 2({\bf 1}',{\bf 2}')_{-5/3},~ 2({\bf 1}',{\bf 2}')_{5/3},\label{EqHS1}\\
&& T_2:~ ({\bf 1}',{\bf 2}')_{5/3},~ ({\bf 1}',{\bf 2}')_{-5/3},\label{EqHS3}\\
&& T_1:~ (\overline{\bf 10}',{\bf 1}')_0 ,~  ({\bf 5}',{\bf 2}')_0
,~  (\overline{\bf 5}',{\bf 1}')_0,~ ({\bf 1}',{\bf 2}')_0,
~(\overline{\bf 5}',{\bf 1}')_{-5/3},~ ({\bf 1}',{\bf 2}')_{-5/3},~ 2({\bf 1}',{\bf 2}')_{5/3},\label{EqHS2}\\
&& T_7:~ ({\bf 5}',{\bf 1}')_{5/3},~ 2({\bf 1}',{\bf 2}')_{-5/3},~ ({\bf 1}',{\bf 2}')_{5/3}
.\label{EqHS4}
\end{eqnarray}
Here, we replaced again the right-handed states in
the $T_5$ sector by the left-handed ones in $T_7$ by ${\cal CTP}$
conjugations. We have not included non-abelian group singlets.
The vector-like representations in the above achieve superheavy masses when the neutral singlet under the flipped-SU(5) develop VEVs of order the string scale. We will discuss it in Sec. \ref{sec:SMsinglets}.
Removing vectorlike representations from Eqs. (\ref{EqHS1}--\ref{EqHS4}), there remain
\begin{equation}
(\overline{\bf 10}',{\bf 1}')_0 ,~  ({\bf 5}',{\bf 2}')_0 ,~  (\overline{\bf 5}',{\bf 1}')_0,~ ({\bf 1}',{\bf 2}')_0.\label{HSremain}
\end{equation}
The hidden sector SU(2)$'$ is broken by a GUT scale VEV of $({\bf 1}',{\bf 2}')_0$ of (\ref{HSremain}).
\begin{table}[t]
{\tiny
\begin{center}
\begin{tabular}{|c|c|c||c|}
\hline  $P+2V$ & $~\chi~$ & $(N^L)_j$  &~SU(5)$_X$~
\\
\hline $\left(0~0~0~0~0~;\frac{-1}{3}~\frac{-1}{3}~\frac{-1}{3}
\right)(0~0~0~0~0~\frac{-1}{2}~\frac{1}{2}~0)'$ & L & $2_{\bar 1},2_{3}$ &  $(1+1)\cdot {\bf 1}_0$
\\
$\left(0~0~0~0~0~;\frac{-1}{3}~\frac{-1}{3}~\frac{-1}{3}
\right)(0~0~0~0~0~\frac{1}{2}~\frac{-1}{2}~0)'$ & L & $2_{\bar 1},2_{3}$ &  $(1+1)\cdot {\bf 1}_0$
\\
\hline \hline
 $P+2V_+$ & $~\chi~$ & $(N^L)_j$ &~(SU(5)$_X$; SU(5)$'$, SU(2)$'$)~
\\
\hline
$\left(\frac{-1}{6} ~\frac{-1}{6} ~\frac{-1}{6} ~\frac{-1}{6} ~\frac{-1}{6}~;
\frac{1}{6} ~\frac{-1}{6} ~\frac12\right)
\left(\frac{1}{3} ~\frac{1}{3} ~\frac{1}{3}~\frac{1}{3} ~0
~\frac{1}{6}~\frac{1}{2}~0\right)'$ & L &
$1_{\bar 1}$ &  $1\cdot ({\bf 1}_{5/3};{\bf 1}', {\bf 2}'$)
\\
$\left(\frac{-1}{6} ~\frac{-1}{6} ~\frac{-1}{6} ~\frac{-1}{6} ~\frac{-1}{6}~;
\frac{1}{6} ~\frac{-1}{6} ~\frac12\right)
\left(\frac{-1}{6} ~\frac{-1}{6} ~\frac{-1}{6}~\frac{-1}{6} ~\frac{-1}{2}
~\frac{-1}{3}~0~\frac{-1}{2}\right)'$ & L &
$1_{\bar 1}$ &
\\ \hline
$\left(\frac{1}{3} ~\frac{1}{3} ~\frac{1}{3} ~\frac{1}{3} ~\frac{1}{3}~;
\frac{-1}{3} ~\frac{1}{3} ~0\right)
\left(\frac{-1}{6} ~\frac{-1}{6} ~\frac{-1}{6}~\frac{-1}{6} ~\frac12
~\frac{-1}{3}~0~\frac12\right)'$ & L &
$0$ &  $1\cdot {\bf 1}_{-10/3}$
\\
$\left(\frac{-1}{6} ~\frac{-1}{6} ~\frac{-1}{6} ~\frac{-1}{6} ~\frac{-1}{6}~;
\frac{1}{6} ~\frac{-1}{6} ~\frac12\right)
\left(\frac{-1}{6} ~\frac{-1}{6} ~\frac{-1}{6}~\frac{-1}{6} ~\frac12
~\frac{2}{3}~0~\frac{-1}{2}\right)'$ & L &
$0$ &  $1\cdot {\bf 1}_{5/3}$
\\
$\left(\frac{-1}{6} ~\frac{-1}{6} ~\frac{-1}{6} ~\frac{-1}{6} ~\frac{-1}{6}~;
\frac{1}{6} ~\frac{-1}{6} ~\frac12\right)
\left(\frac{-1}{6} ~\frac{-1}{6} ~\frac{-1}{6}~\frac{-1}{6} ~\frac12
~\frac{-1}{3}~0~\frac{1}{2}\right)'$ & L &
$1_{3}$ & $1\cdot {\bf 1}_{5/3}$
\\
\hline\hline
 $P+2V_-$ & $~\chi~$ & $(N^L)_j$ &~(SU(5)$_X$; SU(5)$'$, SU(2)$'$)~
\\
\hline
$\left(\frac{1}{6} ~\frac{1}{6} ~\frac{1}{6} ~\frac{1}{6} ~\frac{1}{6}~;
\frac{1}{6} ~\frac{-1}{2} ~\frac{-1}{6}\right)
\left(\frac{-1}{3} ~\frac{-1}{3} ~\frac{-1}{3}~\frac{-1}{3} ~0
~\frac{-1}{6}~\frac{-1}{2}~0\right)'$ & L &
$1_{3}$  & $1\cdot ({\bf 1}_{-5/3};{\bf 1}', {\bf 2}')$
\\
$\left(\frac{1}{6} ~\frac{1}{6} ~\frac{1}{6} ~\frac{1}{6} ~\frac{1}{6}~;
\frac{1}{6} ~\frac{-1}{2} ~\frac{-1}{6}\right)
\left(\frac{1}{6} ~\frac{1}{6} ~\frac{1}{6}~\frac{1}{6} ~\frac{1}{2}
~\frac{1}{3}~0~\frac{1}{2}\right)'$ & L &
$1_{3}$ &
\\ \hline
$\left(\frac{-1}{3} ~\frac{-1}{3} ~\frac{-1}{3} ~\frac{-1}{3} ~\frac{-1}{3}~;
\frac{-1}{3} ~0~\frac{1}{3} \right)
\left(\frac{1}{6} ~\frac{1}{6} ~\frac{1}{6}~\frac{1}{6} ~\frac{-1}{2}
~\frac{1}{3}~0~\frac{-1}{2}\right)'$ & L &
$0$ & $1\cdot {\bf 1}_{10/3}$
\\
$\left(\frac{1}{6} ~\frac{1}{6} ~\frac{1}{6} ~\frac{1}{6} ~\frac{1}{6}~;
\frac{1}{6} ~\frac{-1}{2} ~\frac{-1}{6}\right)
\left(\frac{1}{6} ~\frac{1}{6} ~\frac{1}{6}~\frac{1}{6} ~\frac{-1}{2}
~\frac{-2}{3}~0~\frac{1}{2}\right)'$ & L &
$0$  & $1\cdot {\bf 1}_{-5/3}$
\\
$\left(\frac{1}{6} ~\frac{1}{6} ~\frac{1}{6} ~\frac{1}{6} ~\frac{1}{6}~;
\frac{1}{6} ~\frac{-1}{2} ~\frac{-1}{6}\right)
\left(\frac{1}{6} ~\frac{1}{6} ~\frac{1}{6}~\frac{1}{6} ~\frac{-1}{2}
~\frac{1}{3}~0~\frac{-1}{2}\right)'$ & L &
$1_{\bar 1}$ & $1\cdot {\bf 1}_{-5/3}$
\\
\hline
\end{tabular}
\end{center}
\caption{Chiral matter  states satisfying $\Theta_{\{0,\pm\}}=0$ in the $T_2^{\{0,\pm\}}$
sectors. }\label{tb:T2}
}
\end{table}
Then, out of the representations of (\ref{HSremain}), there remain one hidden sector family of SU(5)$'$
\begin{equation}
\overline{\bf 10}'_0 ,~  {\bf 5}'_0 .\label{HSonefam}
\end{equation}
which is the key toward the DSB with SU(5)$'$ \cite{DSBSU5}.
\begin{table}[h]
{\tiny
\begin{center}
\begin{tabular}{|c|c|c||c|}
\hline  $P+V$ & $~\chi~$ & $(N^L)_j$  & Reprs.
\\
\hline $\left(0~0~0~0~0~;\frac{-1}{6}~\frac{-1}{6}~\frac{-1}{6}
\right)(\underline{-1~0~0~0}~0~\frac{1}{4}~\frac{1}{4}
~\frac12)'$ & L & $0$ &$1\cdot ({\bf 1}_0; \overline{\bf 10}',{\bf 1}')$
\\
$\left(0~0~0~0~0~;\frac{-1}{6}~\frac{-1}{6}~\frac{-1}{6}
\right)(\underline{\frac12~\frac12~\frac{-1}{2}
~\frac{-1}{2}}~\frac12~\frac{-1}{4}~\frac{-1}{4}~0)'$ & L & $0$ &
\\ [0.3em] \hline
$\left(0~0~0~0~0~;\frac{-1}{6}~\frac{-1}{6}~\frac{-1}{6}
\right)(\underline{1~0~0~0}~0~\frac{1}{4}~\frac{1}{4} ~\frac12)'$&
L & $0$ &
\\
$\left(0~0~0~0~0~;\frac{-1}{6}~\frac{-1}{6}~\frac{-1}{6}
\right)(0~0~0~0~0~\frac{-3}{4}~\frac{-3}{4}~
\frac{-1}{2})'$& L & $0$ &  $1\cdot ({\bf 1}_0; {\bf 5}',{\bf 2}')$
\\
$\left(0~0~0~0~0~;\frac{-1}{6}~\frac{-1}{6}~\frac{-1}{6}
\right)(\underline{\frac12~\frac{-1}{2}~\frac{-1}{2}
~\frac{-1}{2}}~\frac{-1}{2}~\frac{-1}{4}~\frac{-1}{4}~0)'$ & L & $0$ &
\\
$\left(0~0~0~0~0~;\frac{-1}{6}~\frac{-1}{6}~\frac{-1}{6}
\right)(\frac12~\frac{1}{2}~\frac{1}{2}~\frac{1}{2}~
\frac{1}{2}~\frac{-1}{4}~\frac{-1}{4}~0)'$ & L & $0$ &
\\ \hline
$\left(0~0~0~0~0~;\frac{-1}{6}~\frac{-1}{6}~\frac{-1}{6}
\right)(\underline{\frac{-1}{2}~\frac{1}{2}~\frac{1}{2}~
\frac{1}{2}}~\frac{-1}{2}~\frac{-1}{4}~\frac{-1}{4}~0)'$ & L & $0$ &$1\cdot ({\bf 1}_0; \overline{\bf 5}',{\bf 1}')$
\\
$\left(0~0~0~0~0~;\frac{-1}{6}~\frac{-1}{6}~\frac{-1}{6}
\right)(0~0~0~0~-1~\frac{1}{4}~\frac{1}{4}~\frac12)'$ & L & $0$ &
\\ \hline
$\left(0~0~0~0~0~;\frac{-1}{6}~\frac{-1}{6}~\frac{-1}{6}
\right)(0~0~0~0~1~\frac{1}{4}~\frac{1}{4}~\frac12)'$ & L & $0$ &$1\cdot ({\bf 1}_0;{\bf 1}', {\bf 2}')$
\\
$\left(0~0~0~0~0~;\frac{-1}{6}~\frac{-1}{6}~\frac{-1}{6}
\right)(\frac{-1}{2}~\frac{-1}{2}~\frac{-1}{2}~
\frac{-1}{2}~\frac{1}{2}~\frac{-1}{4}~\frac{-1}{4}~0)'$ & L & $0$ &
\\ \hline
$\left(0~0~0~0~0~;\frac{-1}{6}~\frac{-1}{6}~\frac{-1}{6}
\right)(0~0~0~0~0~\frac{-3}{4}~\frac{1}{4}~\frac12)'$ & L & $3_{3}$ & $1\cdot {\bf 1}_0$
\\
$\left(0~0~0~0~0~;\frac{-1}{6}~\frac{-1}{6}~\frac{-1}{6}
\right)(0~0~0~0~0~\frac{1}{4}~\frac{-3}{4}~\frac12)'$ & L & $3_{3}$ & $1\cdot {\bf 1}_0$
\\
$\left(0~0~0~0~0~;\frac{-1}{6}~\frac{-1}{6}~\frac{-1}{6}
\right)(0~0~0~0~0~\frac{1}{4}~\frac{1}{4}~\frac{-1}{2})'$ & L & $\{1_{1},1_{3}\}$, $\{2_{3},1_{2}\}$,$6_{3}$
& $(1+1+1)\cdot {\bf 1}_0$
\\
\hline \hline
 $P+V_+$ & $~\chi~$ & $(N^L)_j$ & Reprs.
\\
\hline
$\left(\frac{1}{6} ~\frac{1}{6} ~\frac{1}{6} ~\frac{1}{6} ~\frac{1}{6}~;\frac{1}{3} ~\frac{-1}{3} ~0\right)
\left(\underline{\frac{-5}{6} ~\frac{1}{6} ~\frac{1}{6}~\frac{1}{6}}~\frac12~\frac{1}{12}~
\frac{-1}{4}~0\right)'$ & L &$0$ &  $1\cdot ({\bf 1}_{-5/3}; \overline{\bf 5}',{\bf 1}')$
\\
$\left(\frac{1}{6} ~\frac{1}{6} ~\frac{1}{6} ~\frac{1}{6} ~\frac{1}{6}~;\frac{1}{3} ~\frac{-1}{3} ~0\right)
\left(\frac{-1}{3}~\frac{-1}{3}~\frac{-1}{3}~\frac{-1}{3} ~0~\frac{7}{12}~\frac{1}{4}~\frac12\right)'$ & L &$0$ &
\\ \hline
$\left(\frac{1}{6} ~\frac{1}{6} ~\frac{1}{6} ~\frac{1}{6} ~\frac{1}{6}~;\frac{1}{3} ~\frac{-1}{3} ~0\right)
\left(\frac{-1}{3}~\frac{-1}{3}~\frac{-1}{3}~\frac{-1}{3} ~0~\frac{-5}{12}~\frac{1}{4}~\frac{-1}{2}\right)'$ & L &
$1_{3}$ & $1\cdot ({\bf 1}_{-5/3};{\bf 1}', {\bf 2}')$
\\
$\left(\frac{1}{6} ~\frac{1}{6} ~\frac{1}{6} ~\frac{1}{6} ~\frac{1}{6}~;\frac{1}{3} ~\frac{-1}{3} ~0\right)
\left(\frac{1}{6} ~\frac{1}{6} ~\frac{1}{6}~\frac{1}{6} ~\frac{1}{2}~\frac{1}{12}~\frac{3}{4}~0\right)'$ & L &
$1_{3}$ &
\\ \hline
$\left(\frac{-1}{3} ~\frac{-1}{3} ~\frac{-1}{3} ~\frac{-1}{3} ~\frac{-1}{3}~;\frac{-1}{6} ~\frac{1}{6} ~\frac12\right)\left(\frac{1}{6} ~\frac{1}{6} ~\frac{1}{6}~\frac{1}{6} ~\frac{-1}{2}~\frac{1}{12}~
\frac{-1}{4}~0\right)'$ & L &$1_{3}$ & $1\cdot {\bf 1}_{10/3}$
\\
$\left(\frac{1}{6} ~\frac{1}{6} ~\frac{1}{6} ~\frac{1}{6} ~\frac{1}{6}~;\frac{-2}{3} ~\frac{2}{3} ~0\right)
\left(\frac{1}{6} ~\frac{1}{6} ~\frac{1}{6}~\frac{1}{6} ~\frac{-1}{2}~\frac{1}{12}~\frac{-1}{4}~0\right)'$ & L &
$0$ &  $1\cdot {\bf 1}_{-5/3}$
\\
$\left(\frac{1}{6} ~\frac{1}{6} ~\frac{1}{6} ~\frac{1}{6} ~\frac{1}{6}~;\frac{1}{3} ~\frac{-1}{3} ~0\right)
\left(\frac{1}{6} ~\frac{1}{6} ~\frac{1}{6}~\frac{1}{6} ~\frac{-1}{2}~\frac{1}{12}~\frac{-1}{4}~0\right)'$ & L &
$1_{2},4_{3}$ &  $(1+1)\cdot {\bf 1}_{-5/3}$
\\
\hline\hline
 $P+V_-$ & $~\chi~$ & $(N^L)_j$ & Reprs.
\\
\hline
$\left(\frac{-1}{6} ~\frac{-1}{6} ~\frac{-1}{6} ~\frac{-1}{6} ~\frac{-1}{6}~;
\frac{-2}{3} ~0 ~\frac{-1}{3}\right)
\left(\frac{1}{3} ~\frac{1}{3} ~\frac{1}{3}~\frac{1}{3} ~0
~\frac{-1}{12}~\frac{1}{4}~\frac12\right)'$ & L &
$0$  & $1\cdot ({\bf 1}_{5/3};{\bf 1}', {\bf 2}')$
\\
$\left(\frac{-1}{6} ~\frac{-1}{6} ~\frac{-1}{6} ~\frac{-1}{6} ~\frac{-1}{6}~;
\frac{-2}{3} ~0 ~\frac{-1}{3}\right)
\left(\frac{-1}{6} ~\frac{-1}{6} ~\frac{-1}{6}~\frac{-1}{6} ~\frac{-1}{2}
~\frac{-7}{12}~\frac{-1}{4}~0\right)'$ & L &
$0$ &
\\ \hline
$\left(\frac{-1}{6} ~\frac{-1}{6} ~\frac{-1}{6} ~\frac{-1}{6} ~\frac{-1}{6}~;
\frac{1}{3} ~0 ~\frac{2}{3}\right)
\left(\frac{1}{3} ~\frac{1}{3} ~\frac{1}{3}~\frac{1}{3} ~0
~\frac{-1}{12}~\frac{1}{4}~\frac12\right)'$ & L &
$0$ & $1\cdot ({\bf 1}_{5/3};{\bf 1}', {\bf 2}')$
\\
$\left(\frac{-1}{6} ~\frac{-1}{6} ~\frac{-1}{6} ~\frac{-1}{6} ~\frac{-1}{6}~;
\frac{1}{3} ~0 ~\frac{2}{3}\right)
\left(\frac{-1}{6} ~\frac{-1}{6} ~\frac{-1}{6}~\frac{-1}{6} ~\frac{-1}{2}
~\frac{-7}{12}~\frac{-1}{4}~0\right)'$ & L &
$0$ &
\\ \hline
$\left(\frac{1}{3} ~\frac{1}{3} ~\frac{1}{3} ~\frac{1}{3} ~\frac{1}{3}~;\frac{-1}{6} ~\frac{1}{2} ~\frac{1}{6}\right)\left(\frac{-1}{6} ~\frac{-1}{6} ~\frac{-1}{6}~\frac{-1}{6}~\frac{1}{2}~\frac{5}{12}~\frac{-1}{4}~0\right)'$ & L &$0$  & $1\cdot {\bf 1}_{-10/3}$
\\
$\left(\frac{-1}{6} ~\frac{-1}{6} ~\frac{-1}{6} ~\frac{-1}{6} ~\frac{-1}{6}~;\frac{-2}{3} ~0~\frac{-1}{3} \right)\left(\frac{-1}{6}~\frac{-1}{6}~\frac{-1}{6}~
\frac{-1}{6}~\frac{1}{2}~\frac{5}{12}~\frac{-1}{4}~0
\right)'$ & L &$1_{3}$ & $1\cdot {\bf 1}_{5/3}$
\\
$\left(\frac{-1}{6} ~\frac{-1}{6} ~\frac{-1}{6} ~\frac{-1}{6} ~\frac{-1}{6}~;
\frac{1}{3} ~0~\frac{2}{3} \right)
\left(\frac{-1}{6} ~\frac{-1}{6} ~\frac{-1}{6}~\frac{-1}{6} ~\frac{1}{2}
~\frac{5}{12}~\frac{-1}{4}~0\right)'$ & L &
$1_{3}$  & $1\cdot {\bf 1}_{5/3}$
\\
\hline
\end{tabular}
\end{center}
\caption{Chiral matter  states satisfying $\Theta_{\{0,\pm\}}=0$ in the $T_1^{\{0,\pm\}}$ sectors.}\label{tb:T1}
}
\end{table}
Representations in (\ref{HSonefam}) do not carry any visible
sector quantum numbers and the flipped-SU(5) is not broken by the
DSB in the hidden sector. Our construction of one family SU(5)$'$
with $N_f=0$ or $1$ vector-like pair of $\fivet'$ and $\fivebt'$
does not change the fate of DSB due to the index theorem. But
inclusion of supergravity effects gives a runaway solution at
large values of the dilaton field \cite{KimGcond09}. But the
barrier separation between the SUSY breaking minimum and the
runaway point must be very high. The barrier separation is
controlled by the hidden sector scale.

Finally, in Table \ref{tb:TbA} we list the so-far neglected components of the vectorlike representations of the hidden sector fields carrying nonvanishing hypercharges.
\begin{table}[h]
{\tiny
\begin{center}
\begin{tabular}{|c|c|c||c|}
\hline  $P+5V$ & $~\chi~$ & $(N^L)_j$ &~SU(5)$_X$~
\\
\hline
 $\left(\underline{1~0~0~0~0}~; \frac{1}{6}~\frac{1}{6}~\frac{1}{6}\right)
 (0~0~0~0~0~\frac{1}{4}~\frac{1}{4}~\frac{-1}{2})'$ & R & $0$ & $1\cdot \overline{\bf 5}_{2}^*$
 \\
 $\left(\underline{-1~0~0~0~0}~; \frac{1}{6}~\frac{1}{6}~\frac{1}{6}\right)
 (0~0~0~0~0~\frac{1}{4}~\frac{1}{4}~\frac{-1}{2})'$ & R & $0$  & $1\cdot {\bf 5}_{-2}^*$
 \\ \hline
 $\left(0~0~0~0~0~; \frac{-5}{6}~\frac{1}{6}~\frac{1}{6}\right)
 (0~0~0~0~0~\frac{1}{4}~\frac{1}{4}~\frac{-1}{2})'$ & R & $2_{1}$ &  $1\cdot {\bf 1}_{0}^*$
 \\
 $\left(0~0~0~0~0~; \frac{1}{6}~\frac{-5}{6}~\frac{1}{6}\right)
 (0~0~0~0~0~\frac{1}{4}~\frac{1}{4}~\frac{-1}{2})'$ & R & $2_{1}$ &  $1\cdot {\bf 1}_{0}^*$
 \\
  $\left(0~0~0~0~0~; \frac{1}{6}~\frac{1}{6}~\frac{-5}{6}\right)
 (0~0~0~0~0~\frac{1}{4}~\frac{1}{4}~\frac{-1}{2})'$ & R & $2_{1}$ &  $1\cdot {\bf 1}_{0}^*$
 \\
\hline \hline $P+5V_+$ & $~\chi~$ & $(N^L)_j$
& ~(SU(5)$_X$; SU(5)$'$, SU(2)$'$)~
\\
\hline
$\left(\frac{-1}{6} ~\frac{-1}{6} ~\frac{-1}{6} ~\frac{-1}{6} ~\frac{-1}{6}~;
\frac{-1}{3}~\frac{-2}{3}~0\right)
\left(\frac{1}{3} ~\frac{1}{3} ~\frac{1}{3}~\frac{1}{3} ~0
~\frac{-1}{12}~\frac{1}{4}~\frac12\right)'$ & R &
$0$  & $1\cdot ({\bf 1}_{-5/3}^*;{\bf 1}' , {\bf 2}')$
\\
$\left(\frac{-1}{6} ~\frac{-1}{6} ~\frac{-1}{6} ~\frac{-1}{6} ~\frac{-1}{6}~;
\frac{-1}{3} ~\frac{-2}{3} ~0\right)
\left(\frac{-1}{6} ~\frac{-1}{6} ~\frac{-1}{6}~\frac{-1}{6} ~\frac{-1}{2}
~\frac{-7}{12}~\frac{-1}{4}~0\right)'$ &R &
$0$ &
\\ \hline
$\left(\frac{-1}{6} ~\frac{-1}{6} ~\frac{-1}{6} ~\frac{-1}{6} ~\frac{-1}{6}~;
\frac{2}{3}~\frac{1}{3}~0\right)
\left(\frac{1}{3} ~\frac{1}{3} ~\frac{1}{3}~\frac{1}{3} ~0
~\frac{-1}{12}~\frac{1}{4}~\frac12\right)'$ & R &
$0$  & $1\cdot ( {\bf 1}_{-5/3}^*;{\bf 1}', {\bf 2}')$
\\
$\left(\frac{-1}{6} ~\frac{-1}{6} ~\frac{-1}{6} ~\frac{-1}{6} ~\frac{-1}{6}~;
\frac{2}{3} ~\frac{1}{3} ~0\right)
\left(\frac{-1}{6} ~\frac{-1}{6} ~\frac{-1}{6}~\frac{-1}{6} ~\frac{-1}{2}
~\frac{-7}{12}~\frac{-1}{4}~0\right)'$ & R &
$0$ &
\\ \hline
$\left(\frac{1}{3} ~\frac{1}{3} ~\frac{1}{3} ~\frac{1}{3} ~\frac{1}{3}~;
\frac{1}{6}~\frac{-1}{6}~\frac12\right)
\left(\frac{-1}{6} ~\frac{-1}{6} ~\frac{-1}{6}~\frac{-1}{6} ~\frac12
~\frac{5}{12}~\frac{-1}{4}~0\right)'$ & R &
$0$  & $1\cdot {\bf 1}_{10/3}^*$
\\
$\left(\frac{-1}{6} ~\frac{-1}{6} ~\frac{-1}{6} ~\frac{-1}{6} ~\frac{-1}{6}~;
\frac{-1}{3} ~\frac{-2}{3} ~0\right)
\left(\frac{-1}{6} ~\frac{-1}{6} ~\frac{-1}{6}~\frac{-1}{6} ~\frac{1}{2}
~\frac{5}{12}~\frac{-1}{4}~0\right)'$ & R &
$1_{1}$  & $1\cdot {\bf 1}_{-5/3}^*$
\\
$\left(\frac{-1}{6} ~\frac{-1}{6} ~\frac{-1}{6} ~\frac{-1}{6} ~\frac{-1}{6}~;
\frac{2}{3} ~\frac{1}{3} ~0\right)
\left(\frac{-1}{6} ~\frac{-1}{6} ~\frac{-1}{6}~\frac{-1}{6} ~\frac{1}{2}
~\frac{5}{12}~\frac{-1}{4}~0\right)'$ & R &
$1_{1}$  & $1\cdot {\bf 1}_{-5/3}^*$
\\
\hline \hline $P+5V_-$ & $~\chi~$ & $(N^L)_j$
& ~(SU(5)$_X$; SU(5)$'$, SU(2)$'$)~
\\ \hline
$\left(\frac{1}{6} ~\frac{1}{6} ~\frac{1}{6} ~\frac{1}{6} ~\frac{1}{6}~;
\frac{-1}{3}~0~\frac{1}{3}\right)
\left(\underline{\frac{-5}{6} ~\frac{1}{6} ~\frac{1}{6}~\frac{1}{6}} ~\frac12
~\frac{1}{12}~\frac{-1}{4}~0\right)'$ & R &
$0$  & $1\cdot ({\bf 1}_{5/3}^*; {\bf 5}^{'*},{\bf 1}')$
\\
$\left(\frac{1}{6} ~\frac{1}{6} ~\frac{1}{6} ~\frac{1}{6} ~\frac{1}{6}~;
\frac{-1}{3}~0~\frac{1}{3}\right)
\left(\frac{-1}{3} ~\frac{-1}{3} ~\frac{-1}{3}~\frac{-1}{3} ~0
~\frac{7}{12}~\frac{1}{4}~\frac12\right)'$ & R &
$0$ &
\\ \hline
$\left(\frac{1}{6} ~\frac{1}{6} ~\frac{1}{6} ~\frac{1}{6} ~\frac{1}{6}~;
\frac{-1}{3}~0~\frac{1}{3}\right)
\left(\frac{-1}{3} ~\frac{-1}{3} ~\frac{-1}{3}~\frac{-1}{3} ~0
~\frac{-5}{12}~\frac{1}{4}~\frac{-1}{2}\right)'$ & R &
$1_{1}$ &  $1\cdot ({\bf 1}_{5/3}^*;{\bf 1}', {\bf 2}^{'})$
\\
$\left(\frac{1}{6} ~\frac{1}{6} ~\frac{1}{6} ~\frac{1}{6} ~\frac{1}{6}~;
\frac{-1}{3}~0~\frac{1}{3}\right)
\left(\frac{1}{6} ~\frac{1}{6} ~\frac{1}{6}~\frac{1}{6} ~\frac12
~\frac{1}{12}~\frac{3}{4}~0\right)'$ & R &
$1_{1}$ &
\\ \hline
$\left(\frac{-1}{3} ~\frac{-1}{3} ~\frac{-1}{3} ~\frac{-1}{3}
~\frac{-1}{3}~; \frac{1}{6}~\frac12~\frac{-1}{6}\right)
\left(\frac{1}{6} ~\frac{1}{6} ~\frac{1}{6}~\frac{1}{6}
~\frac{-1}{2} ~\frac{1}{12}~\frac{-1}{4}~0\right)'$ & R & $1_{1}$
& $1\cdot {\bf 1}_{-10/3}^*$
\\
$\left(\frac{1}{6} ~\frac{1}{6} ~\frac{1}{6} ~\frac{1}{6} ~\frac{1}{6}~;
\frac{2}{3}~0~\frac{-2}{3}\right)
\left(\frac{1}{6} ~\frac{1}{6} ~\frac{1}{6}~\frac{1}{6} ~\frac{-1}{2}
~\frac{1}{12}~\frac{-1}{4}~0\right)'$ & R &
$0$ &  $1\cdot {\bf 1}_{5/3}^{*}$
\\
$\left(\frac{1}{6} ~\frac{1}{6} ~\frac{1}{6} ~\frac{1}{6} ~\frac{1}{6}~;
\frac{-1}{3}~0~\frac{1}{3}\right)
\left(\frac{1}{6} ~\frac{1}{6} ~\frac{1}{6}~\frac{1}{6} ~\frac{-1}{2}
~\frac{1}{12}~\frac{-1}{4}~0\right)'$ & R &
$4_{1},1_{\bar 2}$ &  $(1+1)\cdot {\bf 1}_{5/3}^{*}$
\\
\hline
\end{tabular}
\end{center}
\caption{Chiral matter  states from the $T_5^{\{0,\pm\}}$ sectors. All of them are right-handed states. The ${\cal CTP}$ conjugates with the left-handed chirality are provided by the  states in the $T_7$ sector.
}\label{tb:TbA}
}
\end{table}

\subsection{The other vector-like exotic states}

The remaining charged states under the flipped-SU(5)
are the singlets of SU(5)$\times$SU(5)$'\times$SU(2)$'$. They are
listed as follows.
\begin{eqnarray}
&& T_4:~4\cdot{\bf 1}_{-5/3},~ 4\cdot{\bf 1}_{5/3},
\\
&& T_2:~ {\bf 1}_{-10/3},~ 2\cdot{\bf 1}_{5/3},~ {\bf 1}_{10/3},~
2\cdot{\bf 1}_{-5/3},
\\
&& T_1:~ {\bf 1}_{10/3},~ 3\cdot{\bf 1}_{-5/3},~ {\bf 1}_{-10/3},~
2\cdot{\bf 1}_{5/3},
\\
&& T_7:~ {\bf 1}_{10/3},~ 2\cdot{\bf 1}_{-5/3},~ {\bf 1}_{-10/3},~
3\cdot{\bf 1}_{5/3} .
\end{eqnarray}
These are singlet exotics. Since they are also
vector-like under the flipped SU(5), however, they could obtain
superheavy masses, if the needed neutral singlets develop VEVs of order
the string scale. Hence, we can get the same low energy field
spectrum as that of the MSSM. Such vector-like superheavy exotics
could be utilized \cite{BaeKyae} to explain the recently reported high energy cosmic positron excess \cite{PAMELAe,HuhKK}.

\section{Singlets and Yukawa couplings}\label{sec:SMsinglets}

It is necessary to make exotics vectorlike and heavy.
For this purpose, many singlets are required to develop large
VEVs. In Table \ref{tb:singlets}, we list singlet fields.
At least, the following fields are given large VEVs at the string scale,
\begin{eqnarray}
 S_2, ~S_3,~ S_4,~ S_5, ~S_7,~ S_{11},~ S_{12},~ S_{15},~ S_{16},~
 S_{17},~
S_{18},~ S_{21},  ~ S_{22}~ .
\end{eqnarray}
These VEVs are possible through higher dimensional terms in the
superpotential.

\begin{table}[h]
{\tiny
\begin{center}
\begin{tabular}{|c|c|c|c|c||c|}
\hline sectors & singlet states & $~\chi~$ & $(N^L)_j$ &
${\cal P}(f_0)$ & Label \\
\hline\hline

$T_{4}^0$ &
$\left(0~0~0~0~0~;\frac{-2}{3}~\frac{-2}{3}~\frac{-2}{3}
\right)(0^8)'$ & L & $0$ & $3$  & $S_1$ \\
$T_{4}^0$ & $\left(0~0~0~0~0~;\frac{-2}{3}~\frac{1}{3}~\frac{1}{3}
\right)(0^8)'$ & L & $1_{\bar 1},1_{2},1_{3}$ & $2,3,2$
  & $S_2$ \\
$T_{4}^0$ & $\left(0~0~0~0~0~;\frac{1}{3}~\frac{-2}{3}~\frac{1}{3}
\right)(0^8)'$ & L & $1_{\bar 1},1_{2},1_{3}$ & $2,3,2$
  & $S_3$ \\
$T_{4}^0$ & $\left(0~0~0~0~0~;\frac{1}{3}~\frac{1}{3}~\frac{-2}{3}
\right)(0^8)'$ & L & $1_{\bar 1},1_{2},1_{3}$ & $2,3,2$
 & $S_4$ \\
\hline

$T_{6}$ &  $\left(0~0~0~0~0~;0~ 1~0\right)
 \left(0~0~0~0~0~\frac{1}{2}~\frac{-1}{2}~0 \right)'$ &
L & $0$ &  $2$
 & $S_5$ \\
$T_{6}$ &  $\left(0~0~0~0~0~;0~0 ~1\right)
 \left(0~0~0~0~0~\frac{-1}{2}~\frac{1}{2}~0 \right)'$ &
L & $0$ &  $2$  & $S_6$
\\
$T_{6}$ &  $\left(0~0~0~0~0~;0~ -1~0\right)
 \left(0~0~0~0~0~\frac{-1}{2}~\frac{1}{2}~0 \right)'$ &
L & $0$ &  $2$
 & $S_7$ \\
$T_{6}$ &  $\left(0~0~0~0~0~;0~0 -1\right)
 \left(0~0~0~0~0~\frac{1}{2}~\frac{-1}{2}~0 \right)'$ &
L & $0$ & $2$
 & $S_8$
\\
\hline

$T_{3}$ &
$\left(0~0~0~0~0~;\frac{-1}{2}~\frac{-1}{2}~\frac{-1}{2}
 \right)\left(0~0~0~0~0~\frac{3}{4} ~\frac{-1}{4}~ \frac{-1}{2}\right)'$ & L
 & $0$  & $1$
 & $S_9$ \\
$T_{3}$ &  $\left(0~0~0~0~0~;\frac{-1}{2}~\frac{1}{2}~\frac{1}{2}
 \right)\left(0~0~0~0~0~\frac{3}{4} ~\frac{-1}{4}~ \frac{-1}{2}\right)'$ &L
 & $0$  & $1$
 & $S_{10}$ \\
$T_{3}$ &  $\left(0~0~0~0~0~;\frac{1}{2}~\frac{1}{2}~\frac{-1}{2}
 \right)\left(0~0~0~0~0~\frac{-1}{4} ~\frac{3}{4}~ \frac{-1}{2}\right)'$ &L
 & $0$  & $1$
 & $S_{11}$ \\
$T_{3}$ &  $\left(0~0~0~0~0~;\frac{1}{2}~\frac{1}{2}~\frac{-1}{2}
 \right)\left(0~0~0~0~0~\frac{-1}{4} ~\frac{-1}{4}~ \frac{1}{2}\right)'$ &L
 & $1_{1},1_{3}$ &  $2$,$1$
 & $S_{12}$
\\
\hline $T_{9}$ &
$\left(0~0~0~0~0~;\frac{1}{2}~\frac{1}{2}~\frac{1}{2}
 \right)\left(0~0~0~0~0~\frac{-3}{4} ~\frac{1}{4}~ \frac{1}{2}\right)'$ & L
 & $0$  & $1$
 & $S_{13}$ \\
$T_{9}$ &  $\left(0~0~0~0~0~;\frac{1}{2}~\frac{-1}{2}~\frac{-1}{2}
 \right)\left(0~0~0~0~0~\frac{-3}{4} ~\frac{1}{4}~ \frac{1}{2}\right)'$ & L
 & $0$ & $2$
 & $S_{14}$ \\
$T_{9}$ &  $\left(0~0~0~0~0~;\frac{-1}{2}~\frac{-1}{2}~\frac{1}{2}
 \right)\left(0~0~0~0~0~\frac{1}{4} ~\frac{-3}{4}~ \frac{1}{2}\right)'$ & L
 & $0$  & $2$
 & $S_{15}$ \\
$T_{9}$ &  $\left(0~0~0~0~0~;\frac{-1}{2}~\frac{-1}{2}~\frac{1}{2}
 \right)\left(0~0~0~0~0~\frac{1}{4} ~\frac{1}{4}~ \frac{-1}{2}\right)'$ & L
 & $1_{\bar 1}$,$1_{\bar 3}$ &  $1$,$1$
 & $S_{16}$ \\
\hline $T_{2}^0$ &
$\left(0~0~0~0~0~;\frac{-1}{3}~\frac{-1}{3}~\frac{-1}{3}
\right)(0~0~0~0~0~\frac{-1}{2}~\frac{1}{2}~0)'$ & L & $2_{\bar
1},2_{3}$ & $1,1$
 & $S_{17}$ \\
$T_{2}^0$ &
$\left(0~0~0~0~0~;\frac{-1}{3}~\frac{-1}{3}~\frac{-1}{3}
\right)(0~0~0~0~0~\frac{1}{2}~\frac{-1}{2}~0)'$ & L & $2_{\bar
1},2_{3}$ & $1,1$
 & $S_{18}$ \\ \hline

$T_{1}^0$ &
$\left(0~0~0~0~0~;\frac{-1}{6}~\frac{-1}{6}~\frac{-1}{6}
\right)(0~0~0~0~0~\frac{-3}{4}~\frac{1}{4}~\frac12)'$ & L &
$3_{3}$ & $1$
 & $S_{19}$ \\
$T_{1}^0$ &
$\left(0~0~0~0~0~;\frac{-1}{6}~\frac{-1}{6}~\frac{-1}{6}
\right)(0~0~0~0~0~\frac{1}{4}~\frac{-3}{4}~\frac12)'$ & L &
$3_{3}$ & $1$
 & $S_{20}$ \\
$T_{1}^0$ &
$\left(0~0~0~0~0~;\frac{-1}{6}~\frac{-1}{6}~\frac{-1}{6}
\right)(0~0~0~0~0~\frac{1}{4}~\frac{1}{4}~\frac{-1}{2})'$ & L &
$\{1_{1},1_{3}\}$, $\{2_{3},1_{2}\}$,$6_{3}$ & $1,1,1$
 & $S_{21}$ \\
\hline

$T_{7}^0$ &  $\left(0~0~0~0~0~;
\frac{5}{6}~\frac{-1}{6}~\frac{-1}{6}\right)
 (0~0~0~0~0~\frac{-1}{4}~\frac{-1}{4}~\frac{1}{2})'$ & L & $2_{\bar 1}$ &
 $1$
 & $S_{22}$  \\
$T_{7}^0$ &  $\left(0~0~0~0~0~;
\frac{-1}{6}~\frac{5}{6}~\frac{-1}{6}\right)
 (0~0~0~0~0~\frac{-1}{4}~\frac{-1}{4}~\frac{1}{2})'$ & L & $2_{\bar 1}$ &
 $1$
 & $S_{23}$  \\
$T_{7}^0$ &   $\left(0~0~0~0~0~;
\frac{-1}{6}~\frac{-1}{6}~\frac{5}{6}\right)
 (0~0~0~0~0~\frac{-1}{4}~\frac{-1}{4}~\frac{1}{2})'$ & L & $2_{\bar 1}$ &
 $1$
 & $S_{24}$  \\
\hline

\end{tabular}
\end{center}
\caption{Left-handed SU(5)$\times U(1)_X\times$SU(5)$'\times$ SU(2)$'$ singlet states.  The right-handed states in $T_3$ and $T_5$ are converted to the left-handed ones of $T_9$ and $T_7$,
respectively. }\label{tb:singlets} }
\end{table}

\subsection{Conditions}\label{ssec:Yukawa}

Neglecting the oscillator numbers, $H$-momenta of states in various sectors, $H_{\rm mom,0}$ [$\equiv(\tilde s+k\phi+\tilde r_-)$] are assigned as
\begin{align}
&U_1: (-1,0,0),\quad U_2: (0,1,0),\quad U_3:
(0,0,1),\nonumber\\
&\textstyle T_1:(\frac{-7}{12},\frac{4}{12},\frac1{12}),\quad
 T_2:(\frac{-1}{6},\frac46,\frac16),\quad T_3:
 (\frac{-3}{4},0,\frac{1}{4}),\nonumber\\
&\textstyle
 T_4:(\frac{-1}{3},\frac13,\frac13),\quad
\left\{T_5:(\frac{1}{12},\frac{-4}{12},
\frac{-7}{12})\right\},\quad
T_6:(\frac{-1}{2},0,\frac12),\\
&\textstyle T_7:(\frac{-1}{12},\frac{4}{12},\frac{7}{12}),\quad
T_9:(\frac{-1}{4},0,\frac{3}{4}) , \nonumber
\end{align}
from which $T_5$ will not be used since the chiral fields there are right-handed while the other fields are represented as left-handed. With oscillators, the $H$-momentum [$\equiv (R_1,R_2,R_3)$] are
\begin{equation}
(H_{\rm mom})_j= (H_{\rm mom,0})_j  - (N^L)_j + (N^L)_{\bar{j}} ~
,\quad j=1,2,3 . \label{Hmom}
\end{equation}

The superpotential terms by vertex operators should respect the following selection rules~\cite{ChoiKimbk06}:
\begin{itemize}
\item[(a)] Gauge invariance
 \item[(b)] $H$-momentum conservation with $\phi=
  \left(\frac{5}{12}, \frac{4}{12}, \frac{1}{12} \right)$,
\begin{eqnarray}
 \sum_z R_1 (z) = -1 {\rm~ mod~} 12 , \quad \sum_z R_2 (z) = 1 {\rm~ mod~} 3,\quad \sum_z R_3 (z) = 1 {\rm~ mod~} 12,\label{Hconsv}
\end{eqnarray}
where $z(\equiv A,B,C,\dots)$ denotes the index of states participating in a vertex operator. \item[(c)] Space group selection rules:
\begin{eqnarray}
&& \sum_z k(z) = 0 {\rm~ mod~} 12,\label{modinvk} \\
&& \sum_z \left[ km_f \right] (z) = 0 {\rm~ mod~}
3.\label{modinva}
\end{eqnarray}
\end{itemize}
If some singlets obtain string scale VEVs, however, the condition (b) can be merged into Eq. (\ref{modinvk}) in (c). Our strategy is to construct composite singlets (CSs) which have H-momenta, (1~0~0), $(-1~0~0)$, (0~1~0), (0~$-1$~0), (0~0~1), (0~0~$-1$), using only singlets developing VEVs of order at the string scale $M_{\rm string}$. Then, with any integer set $(l~m~n)$, we can attach an appropriate number of CSs such that they make the total $H$-momentum $(-1~1~1)$. Since their VEVs are of order $M_{\rm string}$, the Yukawa couplings multiplied by them are not suppressed.

\subsection{Composite singlets}\label{ssec:CSinglet}

Specifically, let us consider a CS composed of $S_{2}$ with $(N^L)_j=1_{\bar 1}$, $S_{21}$ with $(N^L)_j=\{2_3,1_2\}$, and $S_{22}$ with $(N^L)_j=2_{\bar 1}$ from $T_4^0$, $T_1^0$, and $T_7^0$, respectively. The CS, ``$S_2S_{21}S_{22}$'' fulfills the selection rules (a) and (c) and its H-momentum is calculated as
$\left[\left(\frac{-1}{3}~\frac{1}{3}~\frac{1}{3}\right)
+(1~0~0)\right]
+\left[\left(\frac{-7}{12}~\frac{4}{12}~\frac{1}{12}\right)
+(0~-1~-2)\right]
+\left[\left(\frac{-1}{12}~\frac{4}{12}~\frac{7}{12}
\right)+(2~0~0)\right]
=(2~0~-1)$. The CS composed of $S_3$ with $(N^L)_j=1_{\bar 1}$ or $1_2$ or $1_3$, $S_5$ ($(N^L)_j=0$), and $S_{17}$ ($(N^L)_j=2_3$)
from $T_4^0$, $T_6$, and $T_2^0$, respectively. ``$S_3S_5S_{17}$'' fulfills also (a) and (c) and its H-momentum is given by $(0~1~-1)$, $(-1~0~-1)$, or $(-1~1~-2)$. Similarly, ``$S_{5}S_{7}$'' satisfies ``(a)'' and ``(c)'' and gives the H-momentum of $(-1~0~1)$.  By multiplying properly $S_2S_{21}S_{22}$, $S_3S_5S_{17}$, $S_{5}S_{7}$ (and their higher
powers), thus, one can indeed construct CSs, whose H-momenta are (1~0~0), $(-1~0~0)$, (0~1~0), (0~$-1$~0), (0~0~1), (0~0~$-1$). For instance, $(1~0~0)$ can be obtained from $(2~0~-1)+(-1~0~1)$, namely $(S_2S_{21}S_{22})(S_{5}S_{7})$. $(0~0~1)$ is achieved from $(S_2S_{21}S_{22})(S_{5}S_{7})^2$.

Then we do not have to take care of the ``$H$-momentum
conservation'' in the selection rule  (b) for the superpotential. One
can easily see that all the states in $T_4^+$ and $T_4^-$ achieve
string scale masses by $\langle S_4\rangle$. The states in
$\{T_2^{+},~ T_2^-\}$, $\{T_1^+,~T_7^-\}$ and $\{T_1^-,~T_7^+\}$
pair up to be superheavy by $\langle S_2\rangle$, $\langle
S_3\rangle$, and $\langle S_4\rangle$.  Similarly, the singlet
states in $\{T_1^+,~T_7^-\}$ and $\{T_1^-,~T_7^+\}$ pair up to be
superheavy.

In order to break the flipped-SU(5) to the SM gauge group, we need GUT scale ($\approx$ string scale in our case) VEVs of $\overline{\bf 10}_H$ and ${\bf 10}_H$. We have them from $T_3$ and $T_9$, respectively.  The term ${\bf 10}_H\overline{\bf 10}_H$ and terms with its higher powers are allowed. Thus, SUSY vacua where $\langle \overline{\bf 10}_H\rangle=\langle {\bf 10}_H\rangle\approx M_{\rm string}\approx M_{\rm GUT}$ exist.

We regard a pair of ${\bf 5}_{h}$ and $\overline{\bf 5}_{h}$ in $T_4^0$ as the Higgs fields containing two Higgs doublets of the MSSM. For the missing partner mechanism, we need the couplings $\overline{\bf 10}_H\overline{\bf 10}_H\overline{\bf 5}_h$ and ${\bf 10}_H{\bf
10}_H{\bf 5}_h$. These couplings are allowed in the superpotential by multiplying CSs, $S_{18}S_{11}S_{16}$ and $S_{17}S_{12}S_{15}$, respectively.

The vector-like five-plets appearing in the $T_6$ sector obtain string scale masses. By $\langle S_{21}\rangle$  one pair of five-plets in $T_7^0$ can pair up with one pair of five-plets in $T_4^0$ to be superheavy. The remaining one pair of the five-plets in $T_4^0$, i.e. $\{{\bf 5}_{h}, \overline{\bf 5}_{h}\}$ can get a
mass term (or $\mu$ term) by $S_{17}S_{18}$ and $S_1$. While a VEV $S_{17}S_{18}$ has been assumed, a VEV $S_1$ is not yet assumed. It can be determined by soft terms such that $\mu\equiv \langle S_{17}S_{18}+S_1\rangle\approx m_{3/2}$ as in the next MSSM.

The MSSM matter states in the $T_4^0$ sector couple to the Higgs ${\bf 5}_{h}$, $\overline{\bf 5}_{h}$ in the same sector. Additionally $\langle S_2S_3S_4\rangle$ can be multiplied to suppress the size of the Yukawa couplings. The matter states in the untwisted sector also can couple to them by $S_2$, $S_3$, and $S_4$: $\overline{\bf 10}_{-1}\overline{\bf 10}_{-1}\overline{\bf 5}_{h}\times \langle S_4^2\rangle$, $\overline{\bf 10}_{-1}{\bf 5}_{3}{\bf 5}_{h}\times \langle S_2S_4\rangle$, and ${\bf 1}_{-5}{\bf 5}_{3}\overline{\bf 5}_{h}\times \langle S_2S_3\rangle$. Since there are in total 21 [$=(2+3+2)\times 3$]
states in $S_2$, $S_3$, and $S_4$, they can be utilized to suppress the size of the Yukawa couplings.

\subsection{White dwarf axions and one pair of Higgsino doublets}\label{ssec:WhiteD}

In this subsection, we comment how the needed horizontal symmetry can arise from our heterotic string compactification. But, we will not endeavor to discuss accidental global symmetries arising at some specific vacua \cite{IWKim06,ChoiKS07,ChoiKS09}.
In our previous paper \cite{BaeHuh09}, we introduced  a variant very light axion to enhance the axion-electron coupling. This enhancement was motivated from the unexpected extra energy loss from the white dwarf evolution \cite{Isern08}. It is needed to distinguish families by the quantum numbers of an Abelian horizontal gauge symmetry U(1)$_H$ so that the mixing angles are of ${\cal O}(10^{-1})-{\cal O}(10^{-3})$. The Peccei-Quinn symmetry broken at $\sim 10^{11}$ GeV cannot achieve this goal due to the small mixing $F_a/M_P\sim 10^{-7}$. Let us choose the $H$ direction as
\begin{equation}
H=\frac12(1~1~1~1~1~3~-1~1)(0~0~0~0~0~a~b~c)'
\end{equation}
where
\begin{equation}
b=2a-20,\quad c=\frac32 a-7.
\end{equation}
Then the $H$ quantum numbers of the visible sector quark and Higgs fields are shown below in the square brackets.
\begin{equation}
\begin{array}{l}
U: \tenbt_1 ~[0],\quad T_4: 2~\tenbt_1 ~[0],\quad U: \fivet_{-3} ~[0],\quad T_4: 2~\fivet_{-3} ~[-1],\\
T_4: 2~\fivet_{-2} ~[1],\quad T_4: 2~\fivebt_2 ~[0],\quad T_7: \fivet_{-2} ~[2],\quad\fivebt_2 ~[1]
\end{array}\label{eq:Hqnum}
\end{equation}
which has a $U(1)_H-SU(5)^2$ anomaly. But this anomaly is cancelled by the Green-Schwarz mechanism \cite{GS84}. The $H$ quantum numbers of (\ref{eq:Hqnum}) are minus of those anticipated in Ref. \cite{BaeHuh09}, and hence can act as the needed horizontal gauge symmetry.

As seen in the previous subsection, one pair of quintet and
anti-quintet in $T_7$ are coupled to one pair of quintet and
anti-quintet in $T_4$ via $\langle S_{21}\rangle$, and the
remaining the other pair in $T_4$ was assumed to contain the MSSM
Higgs. In this subsection, we will assume that $\langle
S_{21}\rangle$ and $\langle S_4S_{16}\rangle$ is fine-tuned to be
zero. It is possible because the quantum numbers of $S_{21}$ and
$S_4S_{16}$ are the same.
%
%
Instead we need the following singlet VEVs to remove two pairs of
Higgs quintet and anti-quintet,
\begin{align}
&T_4:~S_1~[-1]\ {\rm and/or}~ S_2~[-1],\nonumber\\
&T_1:~ S_{19}~[-2]\ {\rm and/or}~  S_{20}~[-2].\nonumber
\end{align}
The U(1)$_H$ invariant couplings of the form $T_4T_4T_4$ remove
two pairs of Higgs quintet and anti-quintet of $T_4$. Note that in
the previous subsection $\langle S_1\rangle$ was adjusted to give
a light mass mass term (``$\mu$ term'') of one pair of the quintet
and anti-quintet in $T_4$. The U(1)$_H$ invariant coupling of the
form $T_1T_4T_7$ removes one pair of Higgs quintet and
anti-quintet out of $T_4$ and $T_7$. Thus, the $3\times 3$
Higgsino mass matrix takes the form,
\begin{equation}
\begin{array}{ccc|c}
S_1[-1] &S_1[-1] & 0& \fivet_{-2}^a[1](T_4)\\
S_1[-1] &S_1[-1] &0 & \fivet_{-2}^b[1](T_4)\\
S_{19}[-2]  &S_{19}[-2]  &0 & \fivet_{-2}^c[2](T_7)\\
\hline
\fivebt_2[0](T_4) & \fivebt_2[0](T_4) & \fivebt_2[1](T_7) &\\
\end{array}
\label{eq:MassHiggsino}
\end{equation}
It is obvious that $\fivebt_2[1](T_7)\equiv \fivebt_{-2}^{EW}$ is
massless at this level. If $\langle S_1\rangle=V_1$ and $\langle
S_{19}\rangle=V_2$ and the Yukawa couplings are set to 1, the
matching massless $\fivet_{-2}^{EW}$ is a linear combination of
fives from $T_4$ and $T_7$,
\begin{equation}
\fivet_{-2}^{EW}=\frac{-V_2(\fivet^a+\fivet^b)
+2V_1\fivet^c}{\sqrt{4V_1^2+2V_2^2}}
\end{equation}
where the superscripts $a,b$ and $c$ denote their origins from $T_4$ and $T_7$ as indicated in Eq. (\ref{eq:MassHiggsino}).


\section{Kaluza-Klein spectrum}
The relatively light KK modes ($M_{\rm KK}<
1/\sqrt{\alpha'})$ associated with the relatively large extra
dimensions can arise only in the non-prime orbifolds such as the
${\bf Z}_{12-I}$. It is because KK excitations are possible only
under trivial (untwisted) boundary condition, which leads to $N=2$
(or $N=4$) SUSY spectra. In the ${\bf Z}_{12-I}$ orbifold, for
instance, the boundary conditions associated with the SU(3)
sub-lattice of the 6D compact space in $U$, $T_3$, $T_6$, and
$T_9$ sectors become trivial and allow $N=2$ SUSY sectors
\cite{KimKyaeKK08}.

\begin{figure}[!]
\vskip 0.5cm \resizebox{0.7\columnwidth}{!}
{\includegraphics{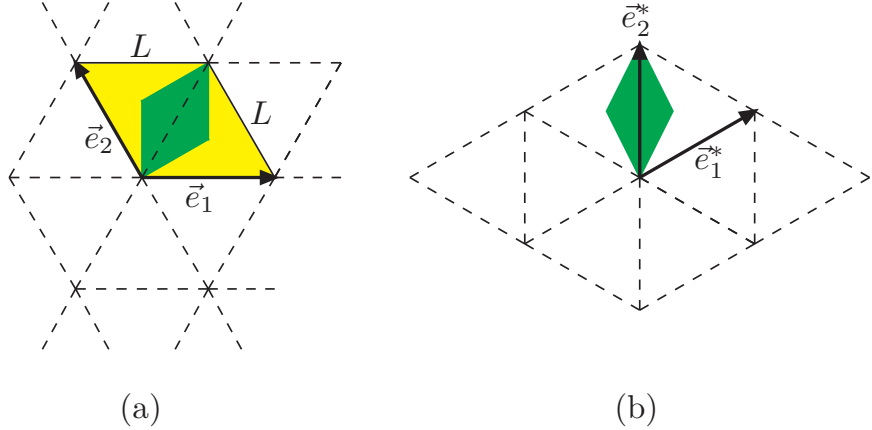}} \caption{The SU(3) lattice (a)
and its dual lattice (b): (a) The torus is inside the yellow
parallelogram and the fundamental region is the green
parallelogram.}\label{fig:Lattice}
\end{figure}

%
The KK modes associated with the relatively large extra dimensions
$R$ ($\equiv R_3=R_4$) of the SU(3) sub-lattice, whose masses
compose KK tower of (integer)/$R$, should also satisfy the
massless conditions \cite{KimKyaeKK08}. Hence, the KK modes in the
$U$ sector still arise from the ${\rm E_8\times E_8}'$ root
vectors\footnote{The states of ${\rm E_8\times E_8}'$ {\it
weights}, not satisfying $P^2=2$, are the string excited sates
with the masses of $({\rm integer})/\sqrt{\alpha'}$}. But $P\cdot
W=$ integer is not necessary for the KK states in
decompactification limit.
%
%
In addition, the GSO projection condition in the U sector is
relaxed from $P\cdot V=$ integer to $P\cdot 3V=$ integer
\cite{KimKyaeKK08}.  The ${\rm E_8\times E_8}'$ roots satisfying
this are
\begin{eqnarray}
{\rm SO(10)} &:& ~\textstyle\left(\underline{\pm 1~\pm
1~0~0~0~};0~0~0\right)(0^8)'
\\
{\rm SO(6)} &:& ~\textstyle\left(0~0~0~0~0~; \underline{\pm 1~\pm
1~0~}\right)(0^8)'
\\
{\rm E_6'} ~~ &:& \left\{
\begin{array}{l}\textstyle ~~ (0^8)\left(
\underline{\pm 1~\pm 1~0~0~0~};~0~0~0\right)'
\\
\textstyle\pm (0^8)\left(\underline{+----}~;+++\right)'
\\\textstyle\pm (0^8)\left(\underline{+++--}~;+++\right)'
\\\textstyle\pm (0^8)\left(+++++~;+++\right)'
\end{array}
\right.
\\
{\rm SU(2)}_{\rm K}' &:& ~\pm(0^8)\left(0~0~0~0~0~;1~-1~0\right)'
.
\end{eqnarray}
Thus, the gauge group is enhanced to
\begin{eqnarray}
{\rm \left[SO(10)\times SO(6)\right]\times\left[E_6\times
SU(2)_{K} \times U(1)\right]'}.
\end{eqnarray}
In the visible sector, the flipped-SU(5) in the massless case are
embedded in a simple group SO(10). Therefore, between the GUT
scale ($\sim$ compactification scale) and string scale, the MSSM
gauge couplings are unified in SO(10), including the U(1)$_X$
coupling. SU(5)$'$ and SU(2)$'$ in the hidden sector are embedded
in E$_6'$. Note that the SU(2)$_{\rm K}'$ emerging
in 6D space is different from the SU(2)$'$ gauge symmetry observed
from the massless spectrum. The SU(2)$'$ is embedded in the
E$_6'$.
The condition for KK matter states ($N=2$ hypermultiplets) from
the $U$ sector is also relaxed from $P\cdot V=$
$\{\frac{-5}{12},~\frac{4}{12},~\frac{1}{12}\}$ (mod Z) to $P\cdot
3V=$ $\pm\frac{1}{4}$ (mod Z) \cite{KimKyaeKK08}.  The KK matter
states from the $U$ sector are shown in TABLE \ref{tb:KKuntwis}.
%
\begin{table} [h]
\begin{center}
\begin{tabular}{|c|c|c|}
\hline  Visible States  & ~4D $\chi~$ & ~SO(10)$\times$SO(6)~\\
\hline  $(~\overline{\bf 16}~;\underline{+--})(0^8)'$ &
 ~L, R~
& $(\overline{\bf 16},{\bf 4})$\\
 $(~\overline{\bf 16}~;+++)(0^8)'$  & L, R & \\
 \hline\hline
Hidden States & ~4D $\chi~$ & ~E$_6'$$\times$SU(2)$_{\rm K}'$~\\
\hline
 $(0^8)(~\overline{\bf 16}~;\underline{+-}-)'$  & L, R &
 \\
$(0^8)(\underline{\pm 1~0~0~0~0}~;\underline{1~0}~0)'$  & L, R &
$(\overline{\bf 27},{\bf 2})'$ \\
 $(0^8)(0~0~0~0~0~;\underline{-1~0}~-1)'$  & L, R &
 \\ \hline
 $(0^8)(0~0~0~0~0~;\underline{-1~0}~1)'$ & L, R & $({\bf 1},{\bf 2})'$\\
\hline
\end{tabular}
\end{center}
\caption{The KK spectrum from the $U$ sector. $\overline{\bf 16}$
collectively denotes $(\underline{+----}$), $(\underline{+++--})$,
and $(+++++$), which are ${\bf 5}$, $\overline{\bf 10}$, and ${\bf
1}$, respectively, in terms of SU(5). Here we drop the ${\cal
CTP}$ conjugates.}\label{tb:KKuntwis}
\end{table}

Among the twisted sectors, only $T_3$, $T_6$ and $T_9$ can provide
KK states in ${\bf Z}_{12-I}$. The KK states from $T_9$ are all
the ${\cal CTP}$ conjugates of the KK states from $T_3$.  As in
the $U$ sector, the KK modes from $T_3$, $T_6$, and $T_9$ should
also satisfy the massless conditions. However, the required GSO
projection is also relaxed.  Following the guide of
Ref.~\cite{KimKyaeKK08}, one can derive the KK spectrum from the
twisted sectors $T_3$ and $T_6$.  The results are presented in
TABLE \ref{tb:KKtwis}.  One can check that the KK spectra in TABLE
\ref{tb:KKuntwis} and \ref{tb:KKtwis} cancel the 6D gauge
anomalies.
%
%
\begin{table} [h]
\begin{center}
\begin{tabular}{|c|c|c|c|c|}
\hline  $P+3V$  & ~$T_k$~ & $(N^L)_j$ & ~4D $\chi~$ & ~SO(10)$\times$SO(6)$\times$SU(2)$_{\rm K}'$~\\
\hline
 $(0^5;\underline{++-})(0^5;\underline{\frac{3}{4}~
 \frac{-1}{4}}~\frac{-1}{2})'$
& $T_3$ & $0$ & L, R &
 $4\times ({\bf 1}, {\bf 4}; {\bf 2}')$\\
  $(0^5;---)(0^5;\underline{\frac{3}{4}~\frac{-1}{4}}~\frac{-1}{2})'$  & $T_3$
  & $0$ & L, R &
\\ \hline
 $(0^5;\underline{++-})(0^5;\frac{-1}{4}~\frac{-1}{4}~\frac{1}{2})'$  & $T_3$
 & $1_1,1_3$ & L, R &
 $8\times ({\bf 1}, {\bf 4}; {\bf 1}')$\\
  $(0^5;---)(0^5;\frac{-1}{4}~\frac{-1}{4}~\frac{1}{2})'$  & $T_3$ &  $1_1,1_3$ & L, R &
 \\ \hline
 $(~\overline{\bf
16}~;0~0~0)(0^5;\frac{-1}{4}~\frac{-1}{4}~\frac{1}{2})'$ & $T_3$ &
$0$ & ~L, R~
& $4\times (\overline{\bf 16},{\bf 1}; {\bf 1}')$\\
 \hline\hline
$P+6V$ & $T_k$ & $(N^L)_j$ & ~4D $\chi~$ & ~SO(10)$\times$SO(6)$\times$SU(2)$_{\rm K}'$~\\
\hline $(0^5;\underline{\pm 1~0~0})(0^5~;\underline{+-}~0)'$  &
$T_6$ & $0$ & L, R &
$3\times ({\bf 1},{\bf 6}; {\bf 2}')$ \\
 $(\underline{\pm 1~0^4}~;0^3)(0^5~;\underline{+-}~0)'$  & $T_6$ & $0$ &  L, R &
 $5\times ({\bf 10},{\bf 1}; {\bf 2}')$ \\
\hline
\end{tabular}
\end{center}
\caption{The KK spectrum from the $T_3$ and $T_6$ sectors. All the
states are the singlets under E$_6'$. $\overline{\bf 16}$ in $T_6$
collectively denotes $(\underline{+----}$), $(\underline{+++--})$,
and $(+++++$), which are ${\bf 5}$, $\overline{\bf 10}$, and ${\bf
1}$, respectively, in terms of SU(5). In the $T_6$ sector, we drop
the ${\cal CTP}$ conjugates.}\label{tb:KKtwis}
\end{table}
The beta function coefficients $b_{\cal G}^{N=2}$ of SO(10) and
E$_6'$ by KK modes with $N=2$ SUSY are
\begin{eqnarray}
b_{\rm SO(10)}^{N=2} &=& -2\times 8 + 2\times (2\times 8+1\times
10) = 36  , \label{betaKKvis}
\\
b_{\rm E_6'}^{N=2} &=& -2\times 12 + 2\times 3\times 2 = -12
\label{betaKKhid}.
\end{eqnarray}

The KK masses are nothing but the excited momenta ($=\vec{m}_3$,
$\vec{m}_4$) in the SU(3) dual lattice in ${\bf Z}_{12-I}$.  The
Wilson line $W^I$ lift some KK spectra and breaks the gauge
symmetry, say ${\cal G}$ to ${\cal H}$. It is because the momentum
vectors $\vec{m}_3$, $\vec{m}_4$ are shifted by $P^IW^I$, where
$P^I$ indicates the E$_8\times$E$_8'$ weight vectors. It is
clearly seen from the expression for KK masses \cite{KimKyaeKK08}:
\begin{eqnarray}
M_{\rm KK}^2=\sum_{m_a,m_b}\frac{2\tilde g^{ab}}{3R^2}\left(m_a-P\cdot
W\right)\left(m_b-P\cdot W\right)
\end{eqnarray}
where $R$ is the radius of the SU(3) torus, $m_a$, $m_b$ ($a,b=3,4$) are integers, and $\tilde g^{ab}$ is defined as,
\begin{eqnarray}
\tilde g^{ab}=
 \left(
\begin{array}{cc}
2 \quad 1
\\
1 \quad 2
\end{array}
\right) .
\end{eqnarray}
%
%
\begin{figure}[!]
\vskip 0.5cm \resizebox{0.5\columnwidth}{!}
{\includegraphics{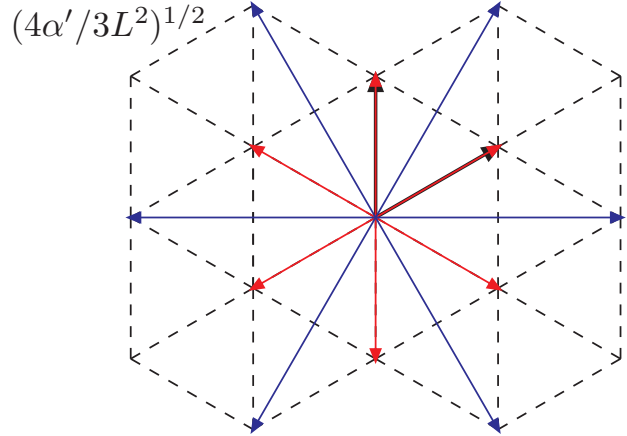}} \caption{The KK modes with $P\cdot
W={\rm integer}$.  The length of the red arrows is
$(4\alpha'/3L^2)^{1/2}$ and that of the blue arrows is
$2(\alpha'/L^2)^{1/2}$.}\label{fig:KKinteger}
\end{figure}
%
%
We list the masses of the first two excited KK states for $P\cdot
W=$ integer:
\begin{eqnarray}
\quad\quad\quad M_{\rm KK}^2 = \left\{
\begin{array}{l}
\frac{4}{3R^2} \quad\quad {\rm for}\quad (m_3,m_4)=\pm(
1,0),~\pm(0,1),~\pm(1,-1) ,
\\
\frac{4}{R^2} \quad\quad {\rm for}\quad
(m_3,m_4)=\pm(1,1),~\pm(2,-1),~\pm(1,-2) .
\end{array}
\right.
\end{eqnarray}
For $P\cdot W=\frac13+$ integer, we have
\begin{eqnarray}
M_{\rm KK}^2 = \left\{
\begin{array}{l}
\frac{4}{9R^2} \quad\quad {\rm for}\quad (m_3,m_4)=(
0,0),~(1,0),~(0,1) ,
\\
\frac{16}{9R^2} \quad\quad {\rm for}\quad
(m_3,m_4)=(1,1),~(1,-1),~(-1,1) .
\end{array}
\right.
\end{eqnarray}
%
\begin{figure}[!]
\vskip 0.5cm \resizebox{0.5\columnwidth}{!}
{\includegraphics{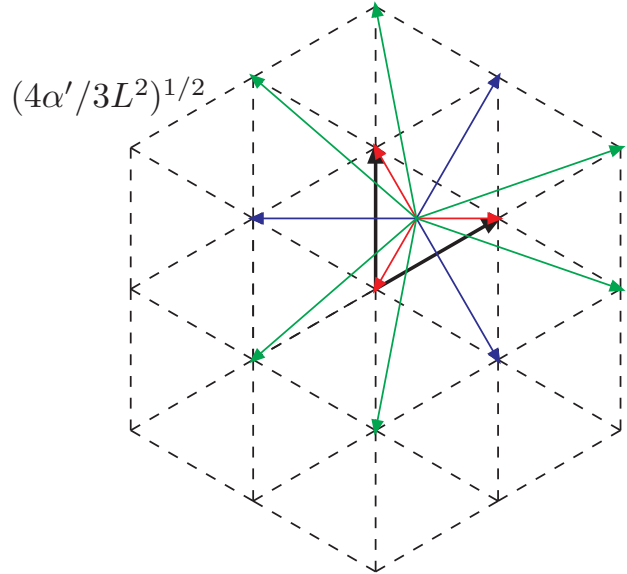}} \caption{The KK modes with $P\cdot
W=\frac13~{\rm mod.~ integer}$. The lengths of the red, blue, and
green arrows are $(\alpha'/L^2)^{1/2}$,  $2(\alpha'/L^2)^{1/2}$,
and $(7\alpha'/L^2)^{1/2}$, respectively.}\label{fig:KKfrac}
\end{figure}
%
In the next excited level, there are 6 KK states, whose
mass-squareds are $\frac{28}{9R^2}$. The KK mass-squareds of the
states with $P\cdot W=-\frac13~+$ integer and $(-m_3,-m_4)$ are
the same as those of the states with $P\cdot W=\frac13~+$ integer
and $(m_3,m_4)$.  Non-vanishing vectors $(m_3,m_4)$ do not affect
the GSO projection conditions.
In TABLE \ref{tb:KKpw}, we display the KK states satisfying
$P\cdot W=$ integer.
\begin{table} [h]
{\tiny
\begin{center}
\begin{tabular}{|c|c|c|c|}
\hline  $P+kV$  & ~$T_k$~ & ~4D $\chi~$ & ~(SU(5),SU(2))(SU(5),SU(2))$'$~\\
\hline  $(\underline{+----}~;+++)(0^8)'$ & $U$ &
 ~L, R~
& $({\bf 5},{\bf 2})({\bf 1},{\bf 1})'$\\
$(\underline{+----}~;+--)(0^8)'$ & $U$ &
 ~L, R~
& \\ \hline
 $(\underline{+++--}~;--+)(0^8)'$  & $U$ & L, R &
 $(\overline{\bf 10},{\bf 1})({\bf 1},{\bf 1})'$ \\
 \hline
$(+++++;-+-)(0^8)'$  & $U$ & L, R & $({\bf 1},{\bf 1})({\bf 1},{\bf 1})'$ \\
 \hline
 $(0^8)(\underline{+---}-~;+--)'$  & $U$ & L, R &
 \\
  $(0^8)(+++++~;+--)'$  & $U$ & L, R & $({\bf 1},{\bf 1})({\bf 5},{\bf 2})'$
 \\
$(0^8)(\underline{1~0~0~0}~0~;1~0~0)'$  & $U$ & L, R &
 \\
 $(0^8)(~0^5~;0~-1~-1)'$  & $U$ & L, R &
 \\
\hline $(0^8)(\underline{+++-}-~;-+-)'$  & $U$ & L, R &
$({\bf 1},{\bf 1})(\overline{\bf 5},{\bf 1})'$ \\
 $(0^8)(~0~0~0~0~-1~;0~1~0)'$  & $U$ & L, R &
 \\
\hline
  $(0^8)(----+~;-+-)'$  & $U$ & L, R & $({\bf 1},{\bf 1})({\bf 1},{\bf 2})'$
 \\
 $(0^8)(0~0~0~0~1~;0~1~0)'$ & $U$ & L, R & \\
\hline $(0^8)(0~0~0~0~0~;0~-1~1)'$ & $U$ & L, R & $({\bf 1},{\bf
1})({\bf 1},{\bf 1})'$
\\
\hline
 $(0^5;-++)(0^5;\frac{3}{4}~
 \frac{-1}{4}~\frac{-1}{2})'$
& $T_3$  & L, R &
 $4\times ({\bf 1}, {\bf 2})({\bf 1},{\bf 1})'$\\
  $(0^5;---)(0^5;\frac{3}{4}~\frac{-1}{4}~\frac{-1}{2})'$  & $T_3$
   & L, R &
\\ \hline
 $(0^5;++-)(0^5;\frac{-1}{4}~
 \frac{3}{4}~\frac{-1}{2})'$
& $T_3$  & L, R &
 $4\times ({\bf 1}, {\bf 1})$\\ \hline
 $(0^5;++-)(0^5;\frac{-1}{4}~\frac{-1}{4}~\frac{1}{2})'$  & $T_3$
  & L, R &
 $8\times ({\bf 1}, {\bf 1})({\bf 1},{\bf 1})'$\\
 \hline
 ~$(~\underline{+++--}~;0~0~0)(0^5;\frac{-1}{4}~\frac{-1}{4}~\frac{1}{2})'$~ & $T_3$ &
 ~L, R~
& $4\times (\overline{\bf 10},{\bf 1})({\bf 1},{\bf 1})'$\\
 \hline $(0^5;0~1~0)(0^5~;+-~0)'$  & $T_6$  & L, R &
$6\times ({\bf 1},{\bf 2})({\bf 1},{\bf 1})'$ \\
 $(0^5;0~0~-1)(0^5~;+-~0)'$  & $T_6$  &  L, R &
 \\ \hline
$(\underline{1~0^4};0^3)(0^5~;-+~0)'$  & $T_6$  & L, R &
$10\times ({\bf 5},{\bf 1})({\bf 1},{\bf 1})'$ \\
\hline
\end{tabular}
\end{center}
\caption{The KK spectrum satisfying $P\cdot W=$ integer.  Here we
drop the ${\cal CTP}$ conjugates.}\label{tb:KKpw} }
\end{table}
Except the states in TABLE \ref{tb:KKpw}, thus, the other KK
states in TABLE \ref{tb:KKuntwis} and \ref{tb:KKtwis} are the
states of $P\cdot W=\pm\frac13~+$ integer.

By the constraint $P\cdot W=$ integer, in the visible sector 6D
SO(10) is broken to the flipped-SU(5), and SO(6) to
SU(2)$\times$U(1)$^2$. The 6D hidden sector gauge group E$_6'$ is
also broken to SU(5)$'\times$SU(2)$'\times$U(1)$'$. But $P\cdot
W=$ integer still leaves intact $N=2$ SUSY.  While the root
vectors of the flipped-SU(5), and SU(5)$'\times$SU(2)$'$ are those
of Eqs.~(\ref{su5root}), (\ref{su5'root}), and (\ref{su2'root}),
the roots of the 6D SU(2) in the visible sector are
$\pm(0^5;0~1~1)(0^8)'$. It is broken to U(1) below the
compactification scale. The beta function coefficient $b_{\cal
H}^{N=2}$ by states of $P\cdot W=$ integer, thus, are
\begin{eqnarray} \label{Bcoeff3}
b_{\rm SU(5)}^{N=2} &=& -2\times 5 + 2\times \left(\frac12\times
12+\frac{3}{2}\times 5\right) = 17  ,
\\ \label{Bcoeff4}
b_{\rm U(1)_X}^{N=2} &=& \frac{1}{40}\times 2\times
\left(3^2\times 10+1^2\times 10+1^2\times 40+2^2\times 50\right) =
17 ,
\\ \label{Bcoeff5}
b_{\rm SU(5)'}^{N=2} &=& -2\times 5 + 2\times \frac12\times 3 = -7
.
\end{eqnarray}
The beta function coefficients $b_{\cal G/H}^{N=2}$ by the
``matter'' states with $P\cdot W=\pm\frac13~+$ integer are
$b_{\cal G}^{N=2}-b_{\cal H}^{N=2}$.
Since $b_{\rm SU(5)}^{N=2}$ is the same as $b_{\rm U(1)_X}^{N=2}$,
and both are included in $b_{\rm SO(10)}^{N=2}$ in
Eq.~(\ref{betaKKvis}), the KK modes in this model do not affect
the gauge coupling unification of SU(5) and U(1)$_X$.
Accordingly, only the fields in $N=1$ SUSY sector, which have no
corresponding KK states, affect the unification.

From the beta function coefficients, we can expect
that the MSSM gauge couplings rapidly increase in the ultraviolet
region. On the other hand, the hidden sector gauge coupling is
asymptotically free. Therefore, a large disparity in the visible
and hidden sector couplings at the compactification scale can be
unified to a single coupling at some scale above the
compactification scale. It is interpreted as the string scale. In
other words, starting with a unified coupling at string scale, the
hidden sector SU(5)$'$ coupling can be of order one at a large
scale.

When a gauge group ${\cal G}$ is broken to a subgroup ${\cal H}$
by Wilson line and further broken to ${\cal H}_0$ by orbifolding
(${\cal H}_0={\cal H}$ in our model), the RG evolution of the
gauge coupling of ${\cal H}_0$, including the effects by KK modes,
is described at low energies by
\begin{eqnarray} \label{RGeq}
\frac{4\pi}{\alpha_{{\cal H}_0}(\mu)} =
\frac{4\pi}{\alpha_*}+b^{N=1}_{{\cal H}_0} {\rm
log}\frac{M_*^2}{\mu^2}+b^{N=2}_{\cal H}\Delta^0 + b^{N=2}_{\cal
G/H }\Delta^\pm .
\end{eqnarray}
We assume that dilaton has been stabilized by a non perturbative
effect \cite{dilatonStab}. It can be discussed also in the context
of SUSY breaking of Ref. \cite{KimGcond09}. In Eq.~(\ref{RGeq}),
$b^{N=2}_{\cal H}\Delta^0$ ($b^{N=2}_{\cal G/H}\Delta^\pm$)
denotes the threshold correction by KK modes of $PW=0$
($\pm\frac13$) mod integer, respecting $N=2$ SUSY.  $b^0_{{\cal
H}_0}$ in Eq.~(\ref{RGeq}) is the beta function coefficient
contributed by $N=1$ SUSY sector states.  As discussed above, the
KK mass towers by the states with $P\cdot W=\frac13~+$ integer and
with $P\cdot W=-\frac13~+$ integer are the same. $b^{N=2}_{\cal
G/H}$ is given by $b^{N=2}_{\cal G}-b^{N=2}_{\cal H}$.

As seen in Eqs.~(\ref{betaKKvis}), (\ref{betaKKhid}), and
(\ref{Bcoeff3}), (\ref{Bcoeff4}), (\ref{Bcoeff5}), the beta
function coefficients by KK modes are quite large. Accordingly,
only the KK states residing in the lowest a few layers in the KK
mass tower would be involved in the RG evolution of the visible
SU(5) gauge coupling, before it reaches ${\cal O}(1)$. So we will
keep only such relatively light KK modes for RG analysis of the
gauge couplings.

If $16/9R^2<M_*^2<28/9R^2$, thus, $\Delta^0$ include the
contributions by 6 KK modes with the mass-squared  $4/3R^2$, while
$\Delta^+$ (and also $\Delta^-$) 3 KK modes of $4/9R^2$ and 3 of
$16/9R^2$. Thus, the threshold corrections by such KK modes are
given by
\begin{eqnarray}
b^{N=2}_{\cal H}\Delta^0 &=& 17\cdot 6\cdot {\rm log}\left(\frac{3R^2M_*^2}{4}\right)
,\\
b^{N=2}_{\cal G/H}\Delta^\pm &=& 19\cdot 3\cdot 2\left[{\rm
log}\left(\frac{9R^2M_*^2}{4}\right)+{\rm log}\left(\frac{9R^2M_*^2}{16}\right)\right] ,
\end{eqnarray}
where ${\cal H}=$SU(5) and ${\cal G}=$SO(10).  We assume
$1/R\approx M_{\rm GUT}$ and $\alpha_*=1$. With $\alpha_{\rm
SU(5)}=\frac{1}{25}$, we estimate $R^2M_*^2\approx
2.5$,\footnote{Considering the first excited KK mass-squared is
$4/9R^2$, one could define the effective compactification scale,
$R_{\rm eff}\equiv \frac32 R$. Then, $R_{\rm eff}^2M_*^2=5.6$. So
at $\mu=M_*/\sqrt{5.6}=0.4\times M_*$, the first excited KK modes
appear.} which is consistent with our assumption
$16/9R^2<M_*^2<28/9R^2$. With  $R^2M_*^2\approx 1.9$, and
\begin{eqnarray}
b^{N=1}_{\rm SU(5)'}=-3\times 5+\frac12\times 3+\frac32 =-12
\end{eqnarray}
by $(\overline{\bf 10}',{\bf 1}')_0$, $({\bf 5}',{\bf 2}')_0$, and
$(\overline{\bf 5}',{\bf 1}')_0$ in Eq.~(\ref{HSonefam}), one can
estimate also the confining scale of the hidden SU(5)$'$. It is
just below $\mu\approx 4/3R\approx 0.8M_*$. Therefore,
e.g. if $M_*=2\times 10^{16}$ GeV, the confining scale of the
hidden sector is $1.6\times 10^{16}$ GeV.  Indeed, the string
scale can be much lowered than $10^{18}$ GeV in the strongly
coupled heterotic string theory (or the heterotic M theory), if
the eleventh space dimension is sizable \cite{heteroticM}.

However, the hidden sector confining scale is very sensitive to
$R^2M_*^2$.  If $M_*^2\lesssim 4/9R^2$, all the KK modes do not contribute
to the RG evolution of the gauge couplings upto the string scale
$M_*$, and so we should adopt only the usual 4D RG equation. If
$M_*=2\times 10^{16}$ GeV and so $\alpha_{\rm SU(5)'}^{-1}=25$ at
that scale, the confining scale can be much lower down to
$10^{11}$ GeV.  Here, we assumed SU(2)$'$ is broken and only
$\overline{\bf 10}'$ and ${\bf 5}'$ draw down the confining
scale.

Below the confinement energy scale, the order parameters are
composite fields rather than SU(5)$'$ gaugino and quarks.   As
noticed in Ref.~\cite{KimGcond09}, gaugino condensation scale or
$N=1$ SUSY breaking scale can be much lower than the confinement
scale.  Let us briefly discuss this issue in the following
section.


\section{The hidden sector supersymmetry breaking}
\label{sec:effChiral}

Now, let us proceed to consider the one family SU(5)$'$ model,
with \ten$'$ and \fiveb$'$ plus $N_f$ copies of \hfive\ and
\hfiveb.
%
\begin{figure}[!]
\vskip 0.5cm
\resizebox{0.6\columnwidth}{!}
{\includegraphics{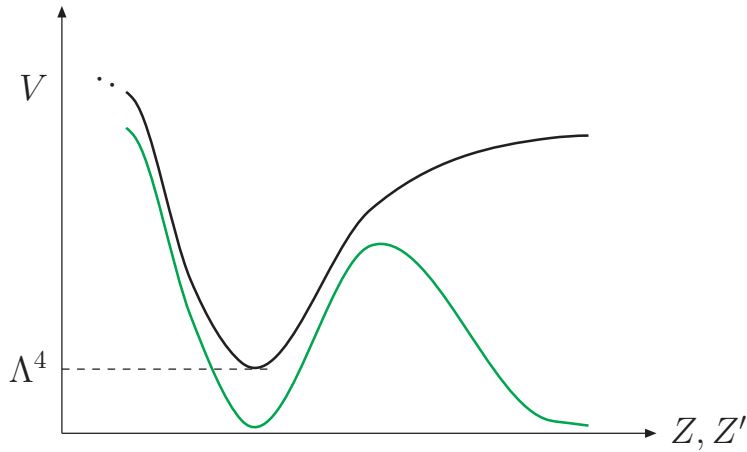}}
\caption{A possible shape of the effective potential in terms of effective fields $Z$ and $Z'$. The lower curve is a schematic view including supergravity effects \cite{KimGcond09}.}\label{fig:EffV}
\end{figure}
%
For $N_f=0$ we can consider two composite chiral fields which are SU(5)$'$ singlets \cite{Meurice84,Kim07st},
\begin{eqnarray}
&&  {\cal W}^\alpha_\beta {\cal W}^\beta_\alpha ,
\nonumber\\
&& \epsilon_{\alpha\gamma\eta\chi\xi} {\cal W}^\alpha_\beta {\cal W}^\gamma_\delta \tent'^{\epsilon\beta}\fivebt'_\epsilon \tent'^{\eta\delta}\tent'^{\chi\xi},\nonumber
\end{eqnarray}
where ${\cal W}^\alpha_\beta$ is the hidden sector gluino superfield, satisfying  ${\cal W}^\alpha_\alpha=0, (\alpha=1,2,\cdots,5)$. There is no more SU(5)$'$-invariant independent chiral combination. For $N_f\ne 0$ also, due to the flavor symmetries of $\fivet'$ and $\fivebt'$, SU$(N_f)\times$SU$(N_f+1)$, we consider only two composite SU(5)$'$ singlet directions affected by instantons \cite{KimNilles09},
\begin{eqnarray}
&& Z\sim {\cal W}^\alpha_\beta {\cal W}^\beta_\alpha ,
\label{Zdef}\\
&& Z'\sim\epsilon_{\alpha\gamma\eta\chi\xi} {\cal W}^\alpha_\beta {\cal W}^\gamma_\delta \tent'^{\nu\beta}\fivebt'_{[\nu } \tent'^{\eta\delta}\tent'^{\chi\xi}(\fivet'^{\mu} \fivebt'_{\mu]}),\label{Zpdef}
\end{eqnarray}
where the lower indices $\nu$ and $\mu$ represent antisymmetric combinations.  In terms of these composite chiral fields, it is known that the confining SUSY theory with one family is known to break SUSY dynamically \cite{KimNilles09,KimGcond09}.
In this $F$-term breaking scenario, we can depict the SUSY
breaking minimum as the local minimum in Fig. \ref{fig:EffV}.  In
the lower curve, we show a schematic view including supergravity
effects \cite{KimGcond09}, which has a runaway piece at large
values of $Z'$.

The dynamically generated effective superpotential, respecting these global symmetries plus the \hfive\ flavor symmetry SU($N_f$) and the \hfiveb\ flavor symmetry SU($N_f+1$), can be written as \cite{KimNilles09}
\begin{equation}
W_{\rm SU(5)}=Z\left[\log\left(\frac{Z^{2-N_f} Z'_\Phi}{\Lambda^{3N_c-2-N_f}} \right) -\alpha\right]\label{eq:SUfW}
\end{equation}
where $\alpha$ is a coupling. It was shown that for $N_f=3$, the SUSY conditions cannot be satisfied and SUSY is dynamically broken \cite{KimNilles09}. Due to the index theorem, for any value of $N_f$, SUSY is dynamically broken, in particular in the SU(5)$'$ theory with one \hten\ and one \hfiveb. The model with the fields of Secs. \ref{sec:USfield} and \ref{sec:TSfield} has
five flipped-SU(5) families. But four of them  carry the exotic
U(1)$_X$ charges. So such four pairs should be assumed to be
superheavy to keep the gauge coupling unification. Nontheless the model still contains the ingredients for the dynamical breaking of SUSY included. Inclusion of supergravity effects has been analyzed by one of us \cite{KimGcond09}.

As discussed in Sec. V, the threshold correction by the KK modes
allows a very wide range of the SU(5)$'$ confinement scale, from
$10^{11}$ GeV to $10^{16}$ GeV.  Moreover, as noticed in
Ref.~\cite{KimGcond09}, the gaugino condensation scale can be
quite low compared to the confinement scale.  Thus, even in the
case where the confinement scale is above $10^{13}$ GeV, one can
obtain $N=1$ SUSY breaking effects in the visible sector of order
$10^{2-3}$ GeV via the gravity mediation.  If the condensation
scale is below the $10^{13}$ GeV, SUSY breaking effects in the
visible sector by the gauge mediation can dominate over those by
the gravity mediation, and here one may resort to the gauge
mediation scenario \cite{Kim07st}.

\section{Conclusion}\label{sec:Conclusion}
We have constructed the flipped-SU(5)$\times$SU(5)$'$
model with three families of the MSSM matter states, based on the
${\bf Z}_{12-I}$ orbifold compactification of the heterotic string
theory.  The flipped-SU(5) breaks down to the SM gauge group by
non-zero VEVS of  ${\bf 10}_H$ and $\overline{\bf 10}_H$. The
doublet/triplet splitting problem is very easily resolved, because
the missing partner mechanism simply works in flipped-SU(5). In
this model, we could obtain ${\rm sin}^2\theta_W=\frac38$ at the
string (or GUT) scale as desired. We have shown that all the extra
states beyond the MSSM field spectrum are vector-like
under the flipped-SU(5) and obtain superheavy masses by VEVs of
some neutral singlets.

In this model, the KK modes do not affect the gauge coupling
unification in the visible sector, because the flipped-SU(5) gauge symmetry is enhanced to the SO(10) gauge symmetry above the compactification scale. On the other hand, they could cause a big difference between the visible and hidden gauge couplings at the compactification scale. Depending on
the size of such disparity between the visible and hidden gauge
couplings at the compactification  scale, a wide range of the confining scale of SU(5)$'$ is possible, $10^{11}$ GeV -- $10^{16}$ GeV. With the
hidden matter $\overline{\bf 10}'$ and ${\bf 5}'$, the gaugino
condensation scale or the $N=1$ SUSY breaking scale can be a few orders
lower than the hidden sector SU(5)$'$ confining scale.

\vskip 0.3cm
\acknowledgments{
 This work is supported in part by the Korea Research Foundation, Grant No. KRF-2005-084-C00001. B.K. is also supported by the BK21 Program of the Ministry of Education, Science and Technology.
}



\end{document}